\documentclass[aps,prb,floats,floatfix,twocolumn]{revtex4}

\usepackage{ifthen}
\usepackage{ifpdf}

\ifpdf
\usepackage{graphicx}
\usepackage{epstopdf}
\else
\usepackage{graphicx}
\usepackage{epsfig}
\fi

\usepackage{latexsym}
\usepackage{amsmath}
\usepackage{amssymb}
\usepackage{bm}
\usepackage{wasysym}

\newcommand{\const}{\mbox{const}}

\newcommand{\eexp}{\mbox{e}^}

\newcommand{\mass}{\mathsf{m}}

\newcommand{\tbox}[1]{\mbox{\tiny #1}}

\newcommand{\amatrix}[1]{\begin{matrix} #1 \end{matrix}}

\newcommand{\be}[1]{\begin{eqnarray}\ifthenelse{#1=-1}{\nonumber}{\ifthenelse{#1=0}{}{\label{e#1}}}}
\newcommand{\ee}{\end{eqnarray}} 

\newcommand{\hide}[1]{\textcolor{red}{[hidden text]}}

\renewcommand{\cite}[1]{[\onlinecite{#1}]}

\begin{document}

\title{Random-matrix modeling of semi-linear response, \\
the generalized variable range hopping picture, \\
and the conductance of mesoscopic rings \\
{\rm\footnotesize [Phys. Rev. B 81, 115464 (2010)]}
}

\author{Alexander Stotland$^{1}$,
Tsampikos Kottos$^{2}$
and Doron Cohen$^{1}$}

\affiliation{
$^{1}$Department of Physics, Ben-Gurion University, Beer-Sheva 84105, Israel \\
$^{2}$Department of Physics, Wesleyan University, Middletown, Connecticut 06459, USA}

%\pacs{03.65.-w}{Quantum mechanics}
%\pacs{05.45.Mt}{Quantum chaos}
%\pacs{73.23.-b}{Mesoscopic systems}

\begin{abstract}
Semi-linear response theory determines 
the absorption coefficient of a driven system 
using a resistor network calculation: 
Each unperturbed energy level of a particle 
in a vibrating trap, or of an electron 
in a mesoscopic ring, is regarded 
as a node ($n$) of the network; 
The transition rates ($w_{mn}$) between the nodes  
are regarded as the elements of a random matrix 
that describes the network.   
If the size-distribution of the connecting elements 
is wide (e.g. log-normal-like rather than Gaussian-like)       
the result for the absorption coefficient
differs enormously from the conventional Kubo 
prediction of linear response theory.
We use a generalized variable range hopping scheme 
for the analysis. In particular we apply 
this approach to obtain practical approximations 
for the conductance of mesoscopic rings. 
In this context Mott's picture of diffusion 
and localization is revisited.      
\end{abstract}

%%%%%%%%%%%%%%%%%%%%%%%%%%%%%%%%%%%%%%%%%%%%%%%%%%%%%%%%%%%%%%%%
%%%%%%%%%%%%%%%%%%%%%%%%%%%%%%%%%%%%%%%%%%%%%%%%%%%%%%%%%%%%%%%%

\maketitle

%%%%%%%%%%%%%%%%%%%%%%%%%%%%%%%%%%%%%%%%%%%%%%%%%%%%%%%%
\section{Introduction}
\label{sec:Introduction}

Semi linear response theory (SLRT) \cite{kbr,bls,slr} provides 
a procedure for the calculation of the absorption coefficient 
of a driven system, assuming that there are well defined transition 
rates $w_{mn}$ between levels ${E_n}$ that are ordered by energy. 
In this context it is helpful to regard $w_{mn}^{-1}$ 
as describing resistors that connect nodes of a network,  
a point of view that has become popular in the related studies   
of variable range hopping (VRH) \cite{mott,miller,AHL,pollak,VRHbook,kbv}.
In the random matrix theory (RMT) framework $w_{mn}$ is a random matrix 
whose construction is inspired by analyzing the statistical 
properties of the Hamiltonian of an actual physical system \cite{wigner,bohigas}.  
Three physical applications have been discussed so far:
{\bf (i)} metallic rings driven by electromotive force \cite{bld}; 
{\bf (ii)} metallic grains driven by low frequency radiation \cite{slr};   
{\bf (iii)} cold atoms that are heated up due to the vibrations of a wall \cite{kbw}.   
It is crucial to observe that depending on the parameters that define the physical model,  
the matrix $w_{mn}$ might be {\em banded} and {\em sparse} \cite{prosen,Fyodo}. 
Consequently, non-trivial results that go beyond linear response theory (LRT)
are obtained.

In order to have a precise mathematical definition 
of the RMT model, let us write the random matrix as  
\be{10}  
w_{mn} \ \ = \ \ \bm{X}_{mn} \times \tilde{B}(E_m{-}E_n)
\ee   
In this expression $\tilde{B}(\omega)$ describes 
the band-profile of the matrix, and $\bm{X}_{mn}$ 
is a random matrix whose entries~$x$ are positive 
uncorrelated random numbers. 
%
% that are characterized by a probability density distribution~$\rho(x)$.
%
If $\log(x)$ is widely distributed over many decades, 
as in the case of log-normal or log-box distribution, 
then we say that the matrix is effectively sparse.
Sparsity means that the majority of elements 
are very small compared with the average value.

\begin{figure}[h!]

\includegraphics[clip, width=0.9\hsize]{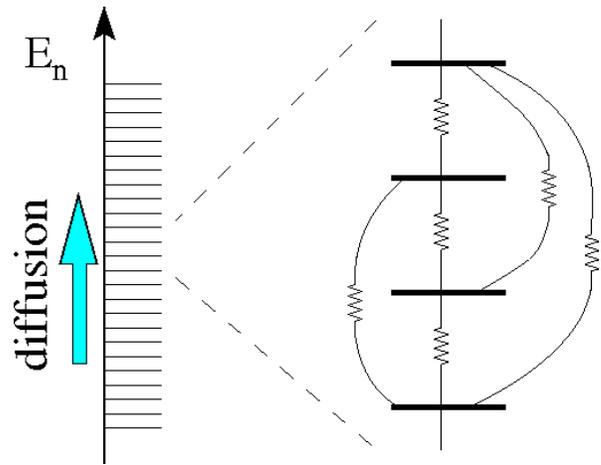}

\caption{
The driving induces transitions between levels $E_n$ of a closed system, 
leading to diffusion in energy space and, hence, an associated heating. 
The diffusion coefficient $D_{\tbox{E}}$ can be calculated using a resistor network analogy. 
Connected sequences of transitions are essential in order 
to have a non-vanishing result, as in the theory of percolation.
}
\label{fig:ResNet}
\end{figure}

Irrespective of real-space dimensionality, 
we regard the index~$n$ of the energy levels  
as labeling the nodes of a 1D lattice (see Fig.\ref{fig:ResNet}), 
hence the $w_{mn}$ define a 1D resistor network. 
The inverse resistivity of this network (see App.~\ref{app:net}) 
is denoted as $w=[[w_{mn}]]$ and has the meaning 
of diffusion coefficient. In proper units the relation is 
\be{12}   
D_{\tbox{E}} 
\ \ &=& \ \ 
\varrho_{\tbox{E}}^{-2} \times [[w_{mn}]] 
\\
\label{e122}
\ \ &\equiv& \ \
\ \ \ \ \ \ G \ \varepsilon^2
\ee
where $\varrho_{\tbox{E}}$ is the mean density of states (DOS).
The parameter $\varepsilon$ represents 
the RMS amplitude of the driving field:
it is the RMS displacement of a wall element 
if we consider the heating of cold atoms in a trap;   
it is the RMS voltage if we consider a ring 
that is driven by an electro-motive-force (EMF). 
We assume here that ${w_{mn}\propto \varepsilon^2}$,  
which holds whenever the standard conditions 
of the Fermi Golden rule (FGR) are satisfied.

Within the framework of the FGR picture
the transitions rates $w_{mn}$ are 
determined by the matrix elements $V_{mn}$
of the perturbation term in the Hamiltonian.  
The naive expectation is to obtain the Kubo formula  
${G = \pi \varrho_{\tbox{E}} \langle\langle |V_{mn}|^2 \rangle\rangle_{\tbox{a}}}$ 
for the absorption coefficient~$G$.
The calculation involves a weighted algebraic average 
\be{536}   
\langle\langle |V_{mn}|^2 \rangle\rangle_{\tbox{a}}
\ \ = \ \ 
\int_{-\infty}^{\infty} 
\langle |V_{mn}|^2 \rangle_{\omega} 
\tilde{F}(\omega)
\frac{d\omega}{2\pi}
\ee
where $\tilde{F}(\omega)$ is a normalized function 
that describes the spectral content of the driving source, 
and $\langle..\rangle_{\omega}$ is defined as the average 
value for $(E_m{-}E_n)\sim\omega$~transitions.   
A more careful inspection reveals that the Kubo 
calculation does not apply to the problem as 
defined above. In order to appreciate the difference 
we re-write Eq.~(\ref{e12}) as 
\be{11}   
G
\ \ = \ \ 
\pi \varrho_{\tbox{E}} 
\langle\langle |V_{mn}|^2 \rangle\rangle 
\ee
The double average notation indicates 
a {\em resistor-network} calculation.  
This SLRT `average' is bounded 
from above by the algebraic average of Eq.~(\ref{e536})
and from below by the corresponding harmonic mean.
Later in this paper we adopt a generalized 
variable range hopping (VRH) procedure 
in order to estimate the SLRT average:
\be{728}   
\langle\langle |V_{mn}|^2 \rangle\rangle
\ \ \approx \ \ 
\int_{-\infty}^{\infty} 
[|V_{mn}|^2]_{\omega} 
\tilde{F}(\omega)
\frac{d\omega}{2\pi}
\ee
where $[..]_{\omega}$ is the typical value 
for $(E_m{-}E_n)\sim\omega$~transitions.    
The notion of typical value will be defined 
later: it is determined by the 
size-distribution of the matrix elements.

Physically the idea behind SLRT is very simple:
in order to have ``good'' absorption it is essential 
to have {\em connected sequences} of transitions. 
Consequently, if $w_{mn}$ is sparse, the traditional 
Kubo expression provides gross over-estimate, because 
it is based on an algebraic average calculation. 
Consequently, our interest is to calculate the {\em ratio} 
between the SLRT and the LRT conductance, 
which we define as the SLRT suppression factor:
\be{0}   
g_{\tbox{SLRT}} 
\ \ \equiv \ \ 
\frac{\langle\langle |V_{mn}|^2 \rangle\rangle}
{\langle\langle |V_{mn}|^2 \rangle\rangle_{\tbox{a}}}
\ee 
where $\langle\langle...\rangle\rangle_{\tbox{a}}$
denotes the usual weighted algebraic average
that appears in the Kubo formula.  
Loosely speaking, if the percentage of {\em large}  
in-band elements is ${s \ll 1}$, 
then a generalized VRH estimate might lead 
to a result of the type 
\be{0}
g_{\tbox{SLRT}} \ \ \sim \ \ \exp\left(-\frac{\const}{s^{\tbox{power}}}\right)         
\ee

A few publications have been devoted to report 
various partial results that have been obtained 
using SLRT. The purpose of the present paper is 
to bridge between SLRT and the traditional literature, 
to further develop the analytical tools, 
and to provide elaborated tangible results 
that hopefully can be tested in actual experiments.

Our main focus concerns the Ohmic conductance $G_{\tbox{Ohm}}$ 
of small metallic rings, which is related to 
the $G$ of Eq.~(\ref{e11}) via ${G_{\tbox{Ohm}}=\varrho_{\tbox{E}}G}$, 
where $E$ is the Fermi energy. Up to a factor, the perturbation 
matrix consists of the elements~$v_{mn}$ of the velocity operator.  
Accordingly, $G_{\tbox{Ohm}}$ is the LRT or the SLRT average over $|v_{mn}|^2$. 
Past literature has provided a theory 
for the conductance in the Debye or adiabatic regimes \cite{rings,wilk} 
where the FGR picture does not apply.  
Diffusive rings have been further analyzed \cite{IS} 
in the Kubo regime, and later weak localization 
corrections have been incorporated \cite{G1,kamenev} 
and verified experimentally~\cite{orsay,expr1,expr2}.

\begin{figure}[h!]
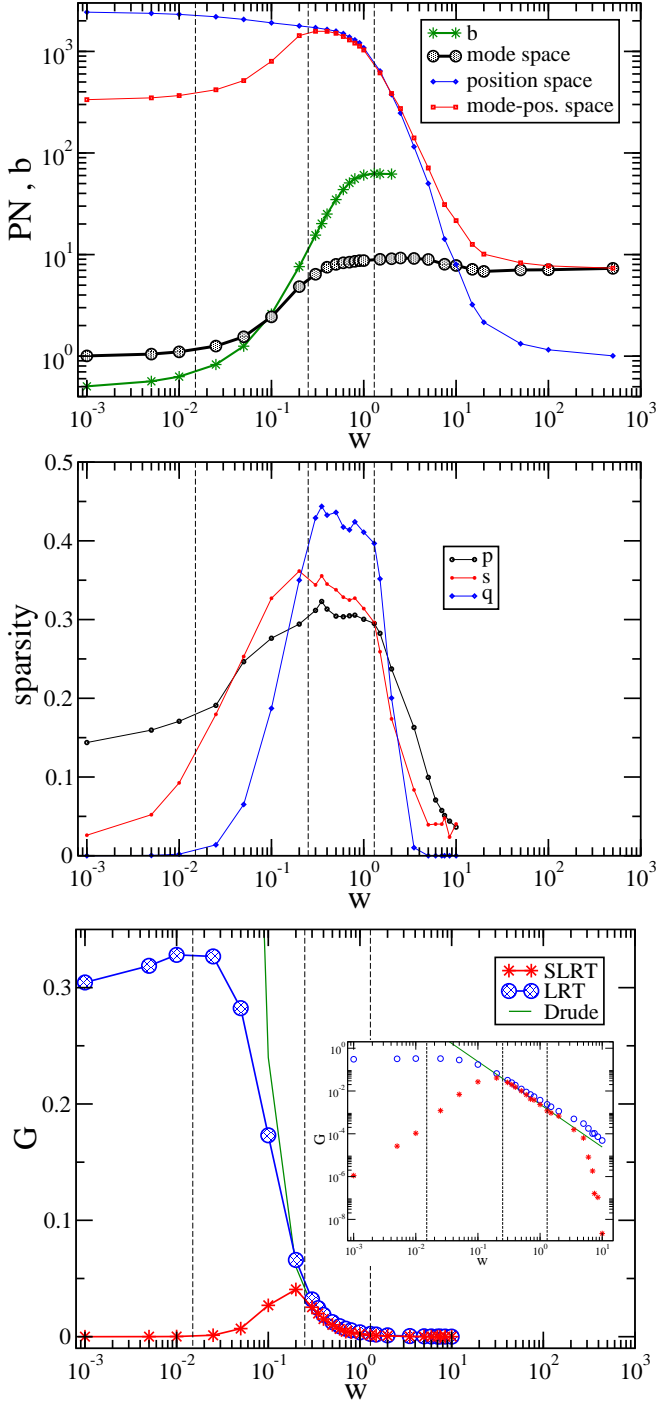


\includegraphics[clip, width=\hsize]{PN}
\includegraphics[clip, width=\hsize]{spq_vs_w}
\setlength{\unitlength}{1mm}
\noindent
\begin{picture}(85,65)
\put(0,0){\includegraphics[clip, width=85mm]{G_vs_W}}
\put(40,18){\includegraphics[clip, width=40mm]{G_vs_W_log}}
\end{picture}

\caption{
{\em Upper panel}: The ergodicity of the eigenstates is characterized 
by the participation number (PN) which is calculated in various 
representations (see App.~\ref{app:model}). 
The bandwidth~$b$ of $v_{nm}$ constitutes another measure for mixing.
The clean, ballistic, diffusive and localization regimes 
(see Sec.~\ref{sec:metallic rings}) are separated by vertical lines.
{\em Middle panel}: The sparsity parameters~($q$, $p$ and $s$) 
that characterize the perturbation matrix $v_{nm}$ 
are plotted versus the disorder~$W$.
{\em Lower panel}: The scaled conductance in arbitrary units 
equals $\langle\langle |v_{mn}|^2 \rangle\rangle$.
The Drude, the LRT and the SLRT results are displayed versus 
the strength of the disorder~$W$. 
{\em Inset}: The same plot in the logarithmic scale.
We see that in the ballistic regime the SLRT conductance 
becomes worse as the disorder becomes weaker, 
in opposition with the Drude expectation.}

\label{fig:measures}
\end{figure}

\begin{figure}[h!]
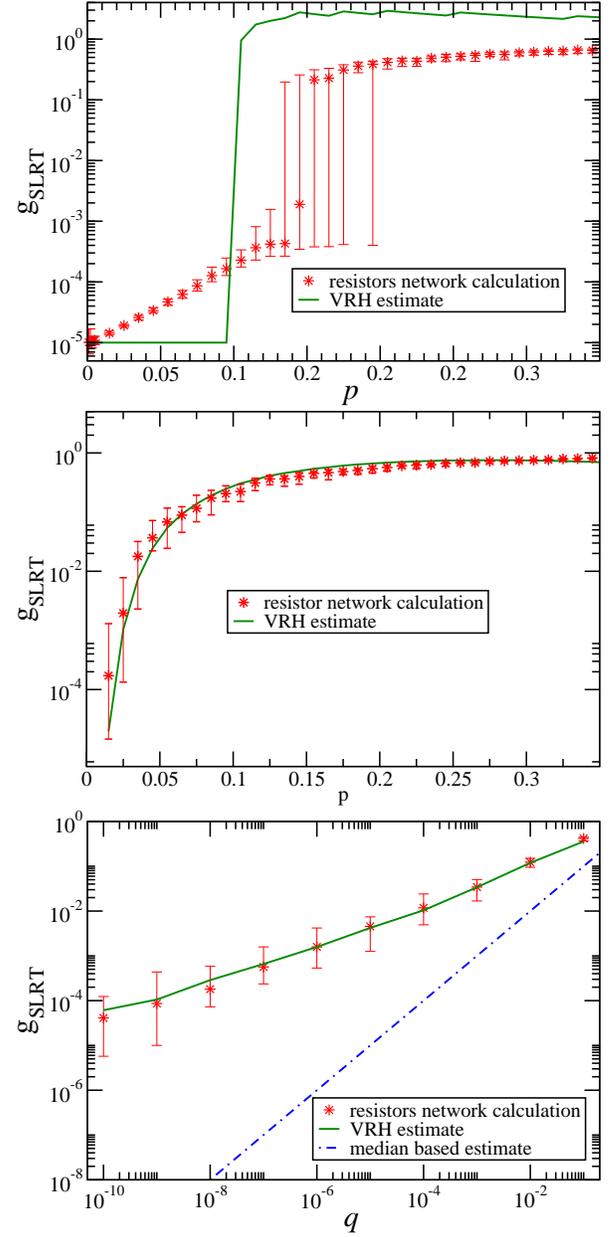


\includegraphics[clip, width=0.9\hsize]{Gmk_RMT_bimodal_rect10_vs_p}
\includegraphics[clip, width=0.9\hsize]{Gmk_RMT_logb_exp10_vs_p}
\includegraphics[clip, width=0.9\hsize]{Gmk_RMT_logn_rect10_vs_q}

\caption{
The SLRT suppression factor $g_{\tbox{SLRT}}$ versus the sparsity 
parameter ($p$ or $q$) for bi-modal distribution with 
rectangular power spectrum (upper panel);
(log-box distribution with exponential power spectrum (middle panel);
log-normal distribution with rectangular power spectrum (lower panel).
The resistor network and the VRH calculation were done for~$100$ 
realizations of ${256 \times 256}$ matrices with ${b=10}$. 
In the big-modal case VRH is not satisfactory. 
In the log-normal case the VRH result is contrasted 
with the naive median based estimate. 
}
\label{fig:G_RMT}
\end{figure}

Still neither VRH in real space, nor SLRT response 
in the ballistic regime had been considered 
in the context of mesoscopic conductance.   
In Fig.~\ref{fig:measures} we present some re-processed 
numerical results that have been reported in Ref.~\cite{bld}. 
These numerical results indicate that indeed 
for both weak and strong disorder 
the matrix elements of the velocity operator become sparse
with ${s\ll1}$. As explained in Ref.~\cite{bld} this is 
related to the non-ergodicity of the eigenstates.
In the present paper we would like to present 
a full analysis of the conductance that starts 
from the strength of the disorder~$W$ as an input. 
The disorder determines the sparsity~$s$, 
and then, using RMT modeling and a generalized VRH approximation, 
leads to some tangible results (Fig.~\ref{fig:G_RMT}) 
for the SLRT suppression factor $g_{\tbox{SLRT}}$. 
We also explain how this factor can be measured in an actual 
laboratory experiment, and how {\em semi-linear} response  
can be distinguished from {\em linear} response in 
a way that does not involve any ambiguities.

{\bf Outline.--} 
{\bf Sec.~\ref{sec:Physical model}} 
motivates the study by introducing the physical model, 
including subsections that relate to the characterization 
of metallic rings and their Kubo-Drude conductance.
Some more details are given in App.~\ref{app:model} and App.~\ref{app:Drude}.
{\bf Sec.~\ref{sec:RMT modeling}} 
discusses the RMT modeling in general.
{\bf Sec.~\ref{sec:The SLRT calculation}} 
briefly reviews the SLRT calculation procedure.  
{\bf Sec.~\ref{sec:VRH approximation}} 
elaborates on the generalized VRH approximation.   
{\bf Sec.~\ref{sec:SLRT analysis}} 
introduces the analysis of some prototype non-Gaussian ensembles. 
Some more details are given in App.~\ref{app:log-box} and App.~\ref{app:log-normal}.
{\bf Sec.~\ref{sec:The semi-classical estimate}} 
discusses the semiclassical theory of the matrix elements 
that are required for the calculation of the mesoscopic conductance.
{\bf Sec.~\ref{sec:RMT ballistic}} 
discusses the SLRT calculation in the ballistic regime.  
{\bf Sec.~\ref{sec:RMT Anderson}} 
discusses the SLRT  calculation in the Anderson localization regime.  
{\bf Sec.~\ref{sec:SLRT vs VRH}} 
clarifies the relation between SLRT and the traditional VRH calculation. 
{\bf Sec.~\ref{sec:VRH from Kubo?}} 
questions the possibility to get VRH from proper LRT analysis.
{\bf Sec.~\ref{sec:VRH vs Hopping}} 
contrasts VRH with non-thermal hopping due to noisy source. 
{\bf Sec.~\ref{sec:experiment}} 
proposes how to experimentally test SLRT via conductance measurements.    
{\bf Sec.~\ref{sec:Summary}} 
summarizes the major observations regarding the relation between SLRT, LRT and
VRH.

%%%%%%%%%%%%%%%%%%%%%%%%%%%%%%%%%%%%%%%%%%%%%%%%%%%%%%%%
\section{Physical model}
\label{sec:Physical model}

In order to physically motivate the analysis, 
we consider a particle of mass $\mass$ 
in a rectangular box of length $L_x=L$ 
and width $L_y$. In one problem, that of Ref.\cite{kbw}, 
we had assumed Dirichlet boundary conditions
and considered the response for vibrations of the wall. 
In the present paper we assume ring geometry 
with periodic boundary conditions on $L_x$, 
and consider the response to electro-motive force (EMF).
In both cases the Hamiltonian matrix can be written as     
\be{7}
\mathcal{H} =  
\mbox{diag}\{ E_{\bm{n}} \} + \{ U_{\bm{m},\bm{n}} \} 
+ f(t) \{V_{\bm{m},\bm{n}}\}
\ee
where $\bm{n}=(n_x,n_y)$ labels the unperturbed 
eigenstates of a clean box/ring, $U(x,y)$ describes 
the potential floor (either smooth deformation or uncorrelated disorder), 
and $V$ is the perturbation matrix due to the driving. 
Given that the energy of the particle is $E$ 
we define $k_{\tbox{E}}=(2\mass E)^{1/2}$ 
and $v_{\tbox{E}}=(2E / \mass)^{1/2}$.  
The associated number of open modes, 
i.e. the number of energetically allowed $n_y$ values, is 
\be{0}
\mathcal{M} \ \ = \ \ \frac{k_{\tbox{E}}L_y}{\pi}
\ee
The density of states is 
\be{23}
\varrho_{\tbox{E}} 
\ \ = \ \  
\frac{\mass}{2\pi} L_xL_y
\ \ = \ \
\mathcal{M}\frac{L}{2v_{\tbox{E}}} 
\ee

The static part of the Hamiltonian can be diagonalized,   
and in the new basis the Hamiltonian takes the form
\be{0}
\mathcal{H} = \mbox{diag}\{ E_n \}  + f(t) \{V_{mn}\}
\ee
where $E_n$ are the perturbed energies.  
The power spectrum $\tilde{S}(\omega)=\varepsilon^2\tilde{F}(\omega)$ 
of the low frequency driving $\dot f$ 
is either rectangular with sharp cutoff 
at some frequency~$\omega_c$, or exponential  
\be{0}
\tilde{F}(\omega) = \frac{1}{2 \omega_c}
\exp\left(-\frac{|\omega|}{\omega_c}\right) 
\ee
We assume that $\omega_c$ is small 
compared with any relevant semi-classical energy scale, 
but larger compared with the mean level spacing.
If the driving is by a thermal source 
then $\omega_c$ can be identified as the {\em temperature} 
of the source. This latter point of view is useful 
in the discussion of the relation between SLRT and VRH.

In the case of an EMF driven ring $\varepsilon$ 
is the RMS of the voltage, and the interaction $-f(t)V$ 
of the particle with the magnetic flux $f(t)$  
involves ${V=-(e/L)v}$, where $v$ is the velocity operator.
Hence 
\be{26}   
V_{mn} \ \ = \ \ \frac{e}{L}v_{mn} \ \ = \ \  \frac{e}{L}(E_m{-}E_n)^2 \ r_{mn}
\ee
where $r$ is the position operator. 
Thus an LRT or an SLRT study of the conductance 
reduces to a study of the statistical properties 
of the so called dipole matrix elements. 
These statistical properties become non-trivial 
for either weak or strong disorder, and they 
should be described by a {\em non-Gaussian ensemble}.  
  
{\em The following subsections contains some extra details 
regarding metallic rings, and can be skipped in first reading.}

%%%%%%%%%%%%%%%%%%%%%%%%%%%%%%%%%%%%%%%%%%%%%%%%%%%%%%%%%%%%%%%%
\subsection{The characterization of metallic rings}
\label{sec:metallic rings}

\begin{figure}[h!]

\includegraphics[clip, width=0.9\hsize]{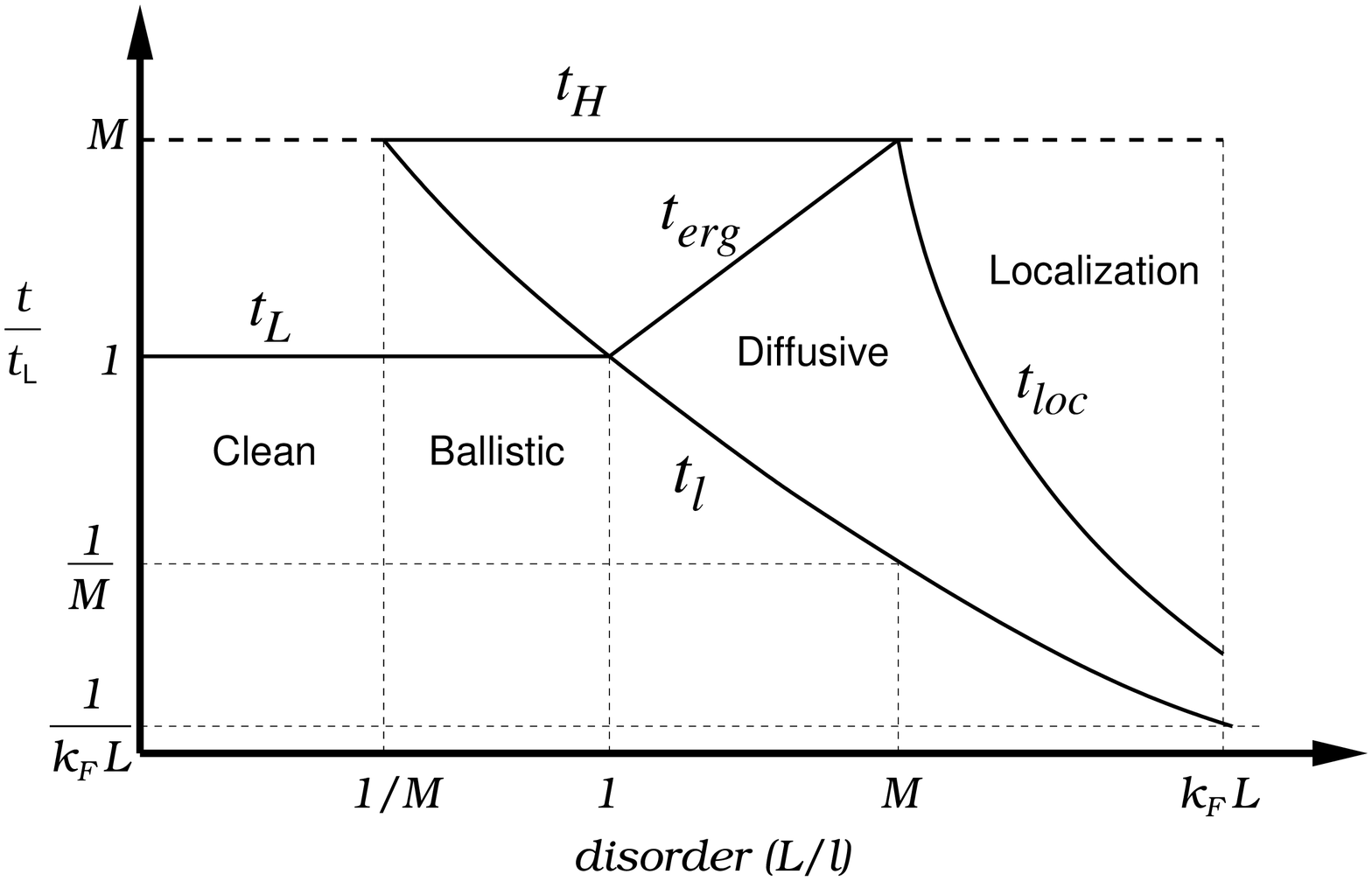} \\
\includegraphics[clip, width=0.9\hsize]{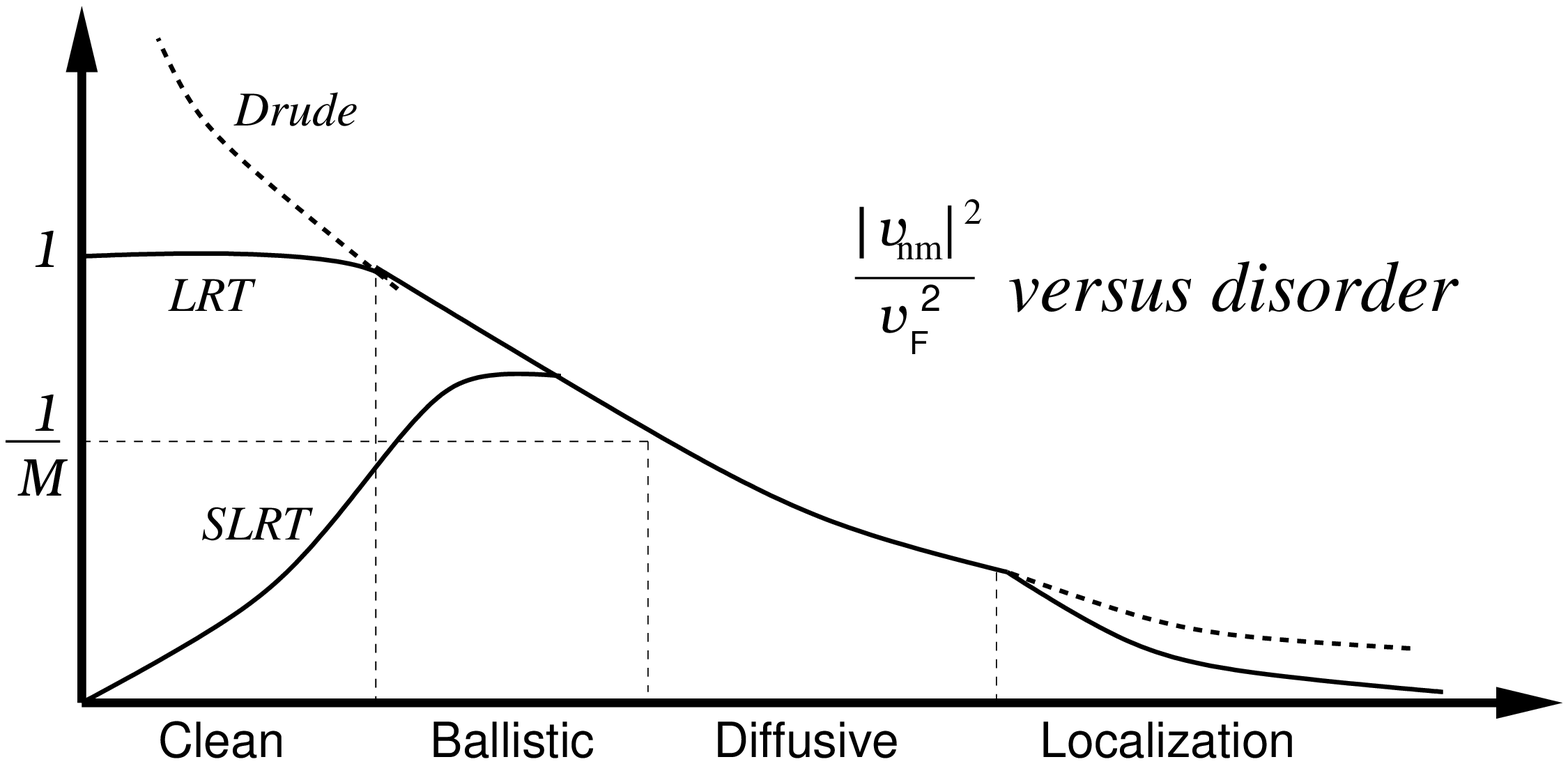}

\caption{
{\em Upper panel:}
Time scales versus the disorder strength (see Sec.\ref{sec:metallic rings}).
{\em Lower panel:}
Schematic illustration that sketch the dependence 
of the DC conductance on the strength of the disorder.
It should be regarded as a caricature of Fig.~\ref{fig:measures}. 
}

\label{fig:time_scales}
\end{figure}

A metallic ring is characterized by the Fermi velocity~$v_{\tbox{E}}$, 
the Fermi momentum~$k_{\tbox{E}}$,   
the length of the ring~$L$, its width~$L_{\perp}$,  
and the strength of the disorder~$W$.  
The latter determines the mean free path $\ell$.
The Fermi velocity $v_{\tbox{E}}$ can be regarded 
as providing conversion between ``length'' and ``time'', 
hence we have two dimensionless parameters:
the number of open modes~$\mathcal{M} \sim (k_{\tbox{E}}L_{\perp})^{d{-}1}$ 
and the degree of disorder $L/\ell$.  
Formally there is a third independent dimensionless 
parameter $k_{\tbox{E}}L$, but we assume it to be very 
large compared with $\mathcal{M}$, 
and hence it has no significant role in the analysis below.  
The various regimes in this problem 
are described below and in the diagram 
of Fig.~\ref{fig:time_scales}:
\begin{itemize}
\setlength{\itemsep}{0mm}
\item Clean ring $(L/\ell) < 1/\mathcal{M}$ 
\item Ballistic ring $(L/\ell) < 1$  
\item Diffusive ring $(L/\ell) > 1$ 
\item Anderson regime $(L/\ell) > \mathcal{M}$. 
\end{itemize}
It is a matter of terminology whether to exclude 
the ``clean'' case from the ballistic regime, 
and the ``Anderson'' case from the ``Diffusive'' regime.

The time scale which is associated with the length 
of the rings is $t_L=L/v_{\tbox{E}}$, 
the time scale which is associated with the 
scattering is mean free time $t_{\ell} = \ell/v_{\tbox{E}}$, 
and the time scale which is associated 
with quantum recurrences 
is the Heisenberg time $t_{\tbox{H}}=\mathcal{M}t_L$.
If the very strong condition ${t_{\ell}> t_{\tbox{H}}}$ 
is satisfied, then we call it ``clean ring'',  
meaning that the disorder does not mix the levels, 
and its effect can be treated 
using first order perturbation theory.

More generally we define the ballistic regime 
by the condition ${\ell \gg L}$. 
If the disorder is strong enough 
then the levels are mixed non-perturbatively 
leading to genuine semi-classical ballistic behavior with
\be{0}      
t_L < t_{\ell} < t_{\tbox{H}}   
\ \ \ \ \ \ \mbox{[Ballistic]}
\ee 
In the diffusive regime it is meaningful     
to define the ergodic (Thouless) time via the 
relation ${D_0t \sim L^2}$ where $D_0=v_{\tbox{E}}\ell$,  
leading to $t_{\tbox{erg}}=(L/\ell)t_L$.
In the strict diffusive regime we have 
\be{0} 
t_{\ell}< t_{\tbox{erg}} < t_{\tbox{H}}
\ \ \ \ \ \ \mbox{[Diffusive]}
\ee
If we have (formally) $t_{\tbox{erg}}>t_{\tbox{H}}$ 
then there is no ergodization but rather 
a strong (Anderson) localization effect shows up. 
This means that one expects a breaktime~$t_{\tbox{loc}}$ 
that marks a crossover from diffusion    
to saturation. A standard argumentation (see below) 
gives the estimate $t_{\tbox{loc}} = \mathcal{M}^2t_{\ell}$.
One observes that in the Anderson regime
\be{0}  
t_{\ell}< t_{\tbox{loc}} < t_{\tbox{H}}
\ \ \ \ \ \ \mbox{[Anderson]}
\ee
The self consistent determination of $\ell_{\xi}$ 
originates in old studies of dynamical 
localization in the quantum kicked rotator problem.  
Assuming that the localization length is $\ell_{\xi}$, 
the local level spacing 
is $\Delta_{\xi}=\pi v_{\tbox{E}}/\mathcal{M}\ell_{\xi}$, 
and hence the breaktime is $t_{\tbox{loc}}=2\pi/\Delta_{\xi}$. 
The self consistency condition 
is $Dt_{\tbox{loc}}\sim \ell_{\xi}^2$, leading to 
\be{0}
\ell_{\xi} \approx \mathcal{M}\ell
\ee
The identification of the Anderson regime is 
via the requirement ${L>\ell_{\xi}}$.  
Finally we note that the largest meaningful  
value of disorder is $(L/\ell)= k_{\tbox{E}}L$,  
for which $\ell$ equals the Fermi wavelength.

%%%%%%%%%%%%%%%%%%%%%%%%%%%%%%%%%%%%%%%%%%%%%%%%%%%%%%%%%%%%%%%%
\subsection{The Kubo-Drude conductance}
\label{sec:The Kubo-Drude conductance}

The coefficient $G$ is defined through the expression ${D_E=G\varepsilon^2}$ 
for the one-particle diffusion coefficient, where $\varepsilon$ 
is the RMS of the voltage. Taking Eq.~(\ref{e26}) into account it follows that 
\be{69}
G = \pi\varrho_{\tbox{E}} \times \left(\frac{e}{L}\right)^2 \ \langle\langle |v_{mn}|^2 \rangle\rangle
\ee 
The Ohmic conductance is defined as the coefficient 
in the Joule formula $G_{\tbox{Ohm}}\epsilon^2$ 
for the rate of energy absorption.  
For an $\mathcal{N}$ particle system at temperature $T$ 
it is related to $G$ via a general 
diffusion-dissipation relation:
\be{0}
G_{\tbox{Ohm}} =& \varrho_{\tbox{E}} \times G 
& \ \ \ \ \mbox{[Fermi]}
\\
G_{\tbox{Ohm}} =& (\mathcal{N}/T) \times G 
& \ \ \ \ \mbox{[Boltzmann]}
\ee 
The Boltzmann occupation applies to semiconductors, 
where $\mathcal{N}/L$ is the density of the particles.
For Fermi occupation $\varrho_{\tbox{E}}/L$ is the density of states 
at the Fermi energy per unit length of the ring, 
and it is in agreement with the Boltzmann result if
we regard $\mathcal{N}=\varrho_{\tbox{E}} T$ as the effective 
number of carriers.

In the case of a diffusive ring it makes sense 
to relate the diffusion in energy to the 
diffusion in real space. This relation holds 
in the strict DC limit. Using the 
Einstein relation $G_{\tbox{Ohm}}=(e/L)^2\varrho_{\tbox{E}}\mathcal{D}$  
we deduce that 
\be{84}   
\mathcal{D} \ \ = \ \ 
\pi\varrho_{\tbox{E}} \times \ \langle\langle |v_{mn}|^2 \rangle\rangle_{\omega_c{\sim}0}
\ee   
It is important to keep in mind that for a disconnected 
ring ${\mathcal{D}=0}$ but still we can get from Eq.~(\ref{e84})
a non-zero result ${G\ne 0}$ because the spectral content 
of the driving may have a finite cut-off frequency~$\omega_c$.

The reference case for all our calculations is the 
Drude result which is obtained for a diffusive ring  
in the semi-classical approximation (see App.~\ref{app:Drude}). 
Assuming a mean free path $\ell$ we write the Drude result as 
\be{85} 
G_{\tbox{Drude}} = \frac{e^2}{2\pi\hbar} \mathcal{M} \ \frac{\ell}{L}
\ee 
where $L$ is the length of the ring, 
and $\mathcal{M}$ is the number of open modes 
(proportional to its cross section).

The quantum Kubo calculation gives in leading order 
the same result as Drude: this is well known, 
and obviously it is also a by product of the 
subsequent analysis. One observes that 
the there is a {\em maximum} Kubo conductance 
which is obtained in the limit of a clean ring, 
i.e. for  $\ell/L=\mathcal{M}$.
For completeness we note that for a ring with transmission~$g_0$,  
the following formal identification applies (see App.~\ref{app:Drude}): 
\be{74}
\frac{\ell}{L} \ \ \Leftrightarrow \ \ \frac{g_0}{1-g_0}
\ \ \ \ \ \ \ \ \ \ \Big[< \mathcal{M}\Big]
\ee 
This makes transparent the relation between the Drude 
and the Landauer results. 
 
In later sections our interest is to find 
the SLRT suppression factor $g_{\tbox{SLRT}}$ 
that determines the ratio $G_{\tbox{Ohm}} / G_{\tbox{Drude}}$. 
For this purpose we have to find not only 
the average value of $|v_{mn}|^2$ but also 
their statistics.

%%%%%%%%%%%%%%%%%%%%%%%%%%%%%%%%%%%%%%%%%%%%%%%%%%%%%%%%%%%%%%%%%%%%%
\section{RMT modeling}
\label{sec:RMT modeling}

Regarded as a random matrix $V_{mn}$ is characterized 
by its band profile, and by the size-distribution
of its elements. The standard RMT modeling due to Wigner 
assumes either full or banded matrix with elements 
that are taken out of a Gaussian distribution. 
But our interest is in circumstances where the size 
distribution is wide, i.e. the elements of $|V_{mn}|^2$  
look like realizations of a random variable~$x$  
whose logarithmic value ($\log(x)$) is distributed 
over several decades.

In practice the $|V_{mn}|^2$  
of a physical model does not have 
an idealized flat band profile. 
Consequently, we write 
\be{31} 
|V_{mn}|^{2}  \ \ = \ \ \bm{X}_{mn} \times \tilde{C}(E_m{-}E_n)
\ee
where $\tilde{C}(\omega)$ describes 
the band-profile of the matrix. 
Numerically the band-profile is obtained 
by averaging separately each diagonal [$(n{-}m)=\const$] 
of the matrix, and plotting the result 
against~$\omega=(E_m{-}E_m) \approx (n{-}m)\varrho_{\tbox{E}}^{-1}$.

The question arises, given a matrix~$A_{mn}$ 
that consist of real non-negative elements, 
how to numerically define its bandwidth $b$, 
its sparsity $s$, and the associated distribution $\rho(x)$ 
of its in-band elements.
For the purpose of this paper it was important to adopt 
an unambiguous definition of~$s$, which 
loosely speaking is defined as the percentage 
of large in-band elements. The suggested procedure  
below is based on the participation number (PN) concept. 
The PN of a set $\{x_i\}$ is defined as 
\be{0}
\mbox{PN} \ \ = \ \ \frac{\left(\sum_i x_i\right)^2}{\sum_i x_i^2}
\ee
and reflects the number of the large elements.   
The procedure to determine~$s$ and $b$ goes as follows: 
{\bf (1)} We consider a truncated $A_{mn}$ 
within the energy window of interest;
{\bf (2)} We calculate the band profile 
by averaging separately the elements over each diagonal;   
{\bf (3)} We construct an untextured  
matrix $A_{mn}^{utx}$ by performing random 
permutations of the elements along the diagonals.
{\bf (4)}  We construct a uniformized  
matrix $A_{mn}^{unf}$ by replacing each of the 
elements of a given diagonal by their average.
{\bf (5)} We calculate the participation 
number of the elements in $A_{mn}$. 
This is like counting the number of large elements.
{\bf (6)} We calculate the participation 
number of the elements in $A_{mn}^{unf}$. 
This is like counting the number of in-band elements. 
{\bf (7)} The ratio of the numbers that have 
been calculated in the previous step is defined 
as the sparsity~$s$. 
{\bf (8)} Likewise the bandwidth~$b$ is 
deduced from the number of in-band elements.

The size distribution $\rho(x)$ refers to 
the in-band elements. 
In order to verify that $A_{mn}$ is really 
like a random matrix we perform the SLRT calculation 
(see Sec.~\ref{sec:The SLRT calculation}) once on $A_{mn}$ and once 
on the untextured matrix $A_{mn}^{utx}$.
If the results are significantly different 
we say that texture, i.e. the non random 
arrangement of the large elements, is important. 
The RMT analysis in this paper assumes 
that texture is not too significant.

In particular we are interested in the 
bi-modal, log-box and log-normal ensembles.
The {\em bi-model distribution} is characterized 
by the probability~$p$ of having a large  
value $x=x_1$, otherwise $x=x_0 \ll x_1$, hence   
\be{0}
\rho(x) = (1-p)\delta(x-x_0) + p\delta(x-x_1) 
\ee
In the case of a  {\em log-box distribution} 
the variable $\ln(x)$ has uniform distribution 
within ${[x_0,x_1]}$, hence 
\be{16}
\rho(x) = \frac{1}{\ln\left(x_1/x_0\right)} \ \frac{1}{x}
\ee
In the case of a {\em log-normal distribution} 
the variable $\ln(x)$ has a Gaussian distribution with 
mean $\ln(x_0)$ and standard deviation~$\sigma$, hence 
\be{17}
\rho(x) = \frac{1}{\sqrt{2 \pi}\sigma}
\ \frac{1}{x}
\eexp{-\frac{(\ln(x/x_0))^2}{2\sigma^2}}
\ee
A random variable can be characterized 
by the algebraic, geometric and harmonic averages
\be{0}
\langle\langle x \rangle\rangle_a &=& \langle x \rangle \\
\langle\langle x \rangle\rangle_g &=& \exp[\langle\log x\rangle] \\
\langle\langle x \rangle\rangle_h &=& [\langle1/x\rangle]^{-1}
\ee
The sparsity of a matrix that consists of 
uncorrelated realizations can be characterized 
by a parameter~$s$ or optionally 
by the parameters~$p$ and~$q$ that 
are defined as follows: 
\be{0}
s \ \ &=& \ \ \langle x \rangle^2 / \langle x^2 \rangle \\
p \ \ &=& \ \ \mbox{Prob}(x {>} \langle x\rangle) \\
q \ \ &=& \ \ \langle\langle x \rangle\rangle_{\tbox{median}}/\langle x \rangle
\ee
By this definition $p$ is the fraction of the elements that 
are larger than the algebraic average and $q$ is the ratio between the median
and the algebraic average.
We regard a matrix as sparse if $s\ll1$  
or equivalently if $p \ll 1$ or $q \ll 1$.

%%%%%%%%%%%%%%%%%%%%%%%%%%%%%%%%%%%%%%%%%%%%%%%%%%%%%%%%
\section{The SLRT calculation}
\label{sec:The SLRT calculation}

As in the standard derivation of the Kubo formula, also within the framework of SLRT, the 
leading mechanism for absorption is assumed to be FGR transitions. 
These are proportional to the squared matrix elements $|V_{mn}|^2$. 
The power spectrum of $\dot{f}(t)$ is $\tilde{S}(\omega)=\varepsilon^2\tilde{F}(\omega)$, 
where $\varepsilon$ is the RMS value of the driving amplitude. 
Consequently, the FGR transition rates are
\be{41}
w_{mn} = 2\pi\frac{|V_{mn}|^2}{(E_m{-}E_n)^2}\tilde{S}(E_m{-}E_n)  
\ee
From Eqs.~(\ref{e10}), (\ref{e31}) and (\ref{e41})
one deduces the identification 
\be{0} 
\tilde{B}(\omega) = \frac{2\pi}{\omega^2} \tilde{C}(\omega) \tilde{S}(\omega)
\ee
The inverse resistivity of the network 
has the meaning of diffusion coefficient, 
and from the definition of $G$ in Eq.~(\ref{e122}) 
we deduce  the {\em SLRT formula}  Eq.~(\ref{e11}) with
\be{0} 
\langle\langle |V_{mn}|^2 \rangle\rangle 
\ \ \equiv \ \
\left[\left[ 
2\varrho_{\tbox{E}}^{-3} \ 
\frac{|V_{mn}|^2}{(E_m{-}E_n)^2} \   
\tilde{F}(E_m{-}E_n)
\right]\right]
\ee
This should be contrasted with the {\em Kubo formula}
that involves an algebraic instead of SLRT average: 
\be{24}   
\langle\langle |V_{mn}|^2 \rangle\rangle_{\tbox{a}}
\equiv
\left[\varrho_{\tbox{E}}^{-1} \sum_{m}
|V_{mn}|^2 
\tilde{F}(E_m{-}E_n)
\right]_{\tbox{avr}}
\ee
with average over the reference state~$n$.  
The average is done over all the states whose 
energy $E_n$ is within the energy window 
of interest. In the metallic context 
it is an average around the Fermi energy.

It is a simple exercise  
to verify  that if all the matrix elements are the same, 
say  ${|V_{mn}|^2 = c_0}$, 
then ${\langle\langle|V_{mn}|^2\rangle\rangle = c_0}$ too. 
Also it is a simple exercise to verify 
that the SLRT formula coincides with the Kubo 
formula if there is no randomness, 
i.e. if $|V_{mn}|^2$ is a well defined function of ${E_m{-}E_n}$. 
But if the matrix is structured or sparse then 
\be{0} 
\langle\langle|V_{mn}|^2\rangle\rangle_{\tbox{h}} 
\ < \ 
\langle\langle|V_{mn}|^2\rangle\rangle
\ \ll \ 
\langle\langle|V_{mn}|^2\rangle\rangle_{\tbox{a}}
\ee
If only neighbouring levels are coupled then 
``adding resistors in series'' (see App.~\ref{app:net}) 
implies equality of the SLRT average 
to the harmonic average: 
\be{46}   
\langle\langle |V_{mn}|^2 \rangle\rangle_{\tbox{h}}
\equiv
\left[\varrho_{\tbox{E}}^{-1}\sum_{m}
\frac{1}{|V_{mn}|^{2}} 
\tilde{F}(E_m{-}E_n)
\right]_{\tbox{avr}}^{-1}
\ee
More generally the harmonic average 
is a gross under-estimate.
A generalized VRH scheme 
that we present in Sec.~\ref{sec:VRH approximation}
provides the following 
approximation for the SLRT average:
\be{47}   
\langle\langle |V_{mn}|^2 \rangle\rangle
\ \ \sim \ \ 
\left[
\varrho_{\tbox{E}}^{-1} \ 
\left[|V_{mn}|^2\right]_{\omega} \   
\ \tilde{F}(\omega)
\right]_{\tbox{max}}
\ee 
where the maximum is calculated with respect to $\omega$.
The typical value $|V_{mn}|^{2}_{\omega}$  
for $\omega$ transitions will be defined precisely 
in Sec.~\ref{sec:VRH approximation}, and it reflects the 
size distribution of the matrix elements.
The VRH integral Eq.~(\ref{e728}) is an ad-hoc 
refinement of Eq.~(\ref{e47}) that 
better interpolates with the LRT result, 
and therefore it is advantageous for actual numerical analysis.

Further analysis (see Sec.~\ref{sec:SLRT analysis}) indicates 
that compared with the weighted harmonic average $\langle\langle|V_{mn}|^2\rangle\rangle_{\tbox{h}}$, 
of Eq.~(\ref{e46}), the corresponding geometric average
$\langle\langle|V_{mn}|^2\rangle\rangle_{\tbox{g}}$ 
provides in most cases a better lower bound.

%%%%%%%%%%%%%%%%%%%%%%%%%%%%%%%%%%%%%%%%%%%%%%%%%%%%%%%%
\section{The generalized VRH approximation}
\label{sec:VRH approximation}

A 1D network is characterized by its  
inverse resistivity $w=[[w_{mn}]]$.  
Inspired by Ref.\cite{AHL}, 
the inverse resistivity can be estimated 
analytically by finding the maximum  
threshold such that the elements ${w_{mn}>w_0}$ 
form a {\em connected cluster}. 
This leads in the present context 
to a generalized VRH estimate which 
we explain in the following paragraph.

Given a threshold~$w$ and truncating 
the bandwidth at~$\omega$, a sufficient 
condition for having a connected cluster 
is to have at least one non-zero 
element per $\omega$~segment:    
\be{0} 
\varrho_{\tbox{E}}\omega \times 
\mbox{Prob}\Big[x\tilde{B}(\omega){>}w\Big] 
\ \ \mbox{larger than unity}
\ee   
We define the typical value $x_{\omega}$ 
for range~$\omega$ transitions via the relation 
\be{0} 
\varrho_{\tbox{E}}\omega \times 
\mbox{Prob}(x{>}x_{\omega}) 
\ \ \sim \ \ 1 
\ee   
and rewrite the condition for having a connected cluster 
in the following suggestive form:  
\be{0} 
w \ \ < \ \ x_{\omega}\tilde{B}(\omega)
\ee   
Thus an under-estimate for the diffusion 
coefficient is $D \sim w \omega^2$   
based on hopping rate $w$ with steps $\omega$. 
The VRH estimate is based on the idea 
to optimize this under-estimate
with respect to $\omega$ and~$w$, leading to 
\be{0} 
D_{\tbox{E}} 
\ \ \sim \ \ 
\Big[\omega^2 x_{\omega}\tilde{B}(\omega)\Big]_{\tbox{max}}
\ee   
where the maximum is with respect to the hopping range~$\omega$.
In the FGR context this leads to Eq.~(\ref{e47})   
with 
\be{0} 
\left[|V_{mn}|^2\right]_{\omega} \ \ \equiv \ \ x_{\omega} \tilde{C}(\omega)
\ee

%%%%%%%%%%%%%%%%%%%%%%%%%%%%%%%%%%%%%%%%%%%%%%%%%%%%%%%%%%%%%%%%
%%%%%%%%%%%%%%%%%%%%%%%%%%%%%%%%%%%%%%%%%%%%%%%%%%%%%%%%%%%%%%%%
\section{SLRT analysis of some prototype non-Gaussian ensembles}
\label{sec:SLRT analysis}

In this section we derive results for SLRT suppression factor $g_{\tbox{SLRT}}$ 
for the bi-modal, for the log-box, and for the log-normal ensembles. 
The bi-modal distribution is the simplest for pedagogical purpose, 
while the log-box and log-normal ensembles are of greater physical relevance.  
The main results are summarized below, while further details 
of the calculation are given in App.~\ref{app:log-box} and App.~\ref{app:log-normal}. 
Fig.~\ref{fig:G_RMT} presents the outcome of numerical analysis 
that tests the accuracy of the generalized VRH approximation.
In later sections we shall see that the presented results are of relevance  
to the study of conductance in the limits of strong and weak disorder.

{\bf The bimodal ensemble.-- } 
In this case there is a minority 
of large elements (${x=x_1}$) 
that have percentage ${p\ll1}$, 
and a majority of small elements (${x=x_0\ll1}$) 
that have percentage ${1{-}p}$.
Consequently, the typical value for $\omega$ transition   
has a percolation-like crossover from~${x_{\omega}=x_0}$ 
to~${x_{\omega}=x_1}$ at the frequency ${\omega=(\varrho_{\tbox{E}}p)^{-1}}$. 
Therefore, 
\be{0}
g_{\tbox{SLRT}} \ \ \sim \ \ 
\begin{cases}
 q \ 
\tilde{F}(\omega{\sim}0),
& bp < 1 
\\
 (1/p) \ 
\tilde{F}(1/(\varrho_{\tbox{E}}p)),
& bp > 1 
\end{cases}
\ee
where $b=\varrho_{\tbox{E}}\omega_c$ is
the dimensionless bandwidth, and ${q \approx x_0/(px_1)}$.
The first expression reflects 
the possibility of {\em majority dominance} 
of the small elements, 
while the second expression reflects 
the possibility of {\em minority dominance}
of the large elements.
Note that the VRH approximately implicitly 
assumes that $F(\omega)$ drops (say) exponentially 
such that ${g_{\tbox{SLRT}} \ll 1}$, 
otherwise the result cannot be trusted.

{\bf The log-box ensemble.-- } 
In this case the probability distribution 
of $\log(x)$ is uniform over many decades. 
Therefore, it is reasonable to assume that the 
result for $g_{\tbox{SLRT}}$ is {\em minority dominated}. 
It is natural to characterize the log-box distribution of Eq.~(\ref{e16})
by a parameter~${\tilde{p} = \left(\ln (x_1/x_0)\right)^{-1}}$, 
and to realize that the percentage of large 
elements is ${p \approx -\tilde{p}\ln \tilde{p}}$. 
Note that the corresponding sparsity parameter is $s \approx 2\tilde{p}$.
The typical value for $\omega$ transitions is  
\be{0}
x_{\omega} \ \ \approx \ \ \frac{1}{\tilde p}
\exp\left(-\frac{1}{\tilde p\varrho_{\tbox{E}}\omega}\right) 
\ \langle\langle x\rangle\rangle_a
\ee
and the VRH estimate, assuming an exponential bandprofile gives   
\be{63}
g_{\tbox{SLRT}} \ \ \sim \ \ 
\frac{1}{\tilde p} 
\ \exp\left[
-2 \left(\frac{1}{\tilde p \ b}\right)^{1/2}
\right]
\ee
Note the similarity, as well as the  
subtle difference, compared with the  
bimodal minority dominance expectation.

{\bf The log-normal ensemble.-- }
In this case the probability distribution 
of $\log(x)$ is a Gaussian centered around the median.
Therefore,  it is reasonable to assume that the 
result for $g_{\tbox{SLRT}}$ is {\em majority dominated}.
It is natural to characterize the log-normal distribution 
of Eq.~(\ref{e17}) by a parameter~$q$,  
which is defined as the ratio of the median to the algebraic average. 
Note that the corresponding sparsity parameter is ${s=q^2}$.
The VRH calculation gives the result  
\be{65}
g_{\tbox{SLRT}} \ \ \sim \ \ 
q \ \exp
\left[ 
\left(\mbox{factor}\times
\ln\left(\frac{1}{q}\right) 
\ln\left(b\right) 
\right)^{1/2}
\right]
\ee
where the factor is determined by the 
bandprofile (it is~$2$~for an exponential bandprofile 
and~$4$ for a rectangular bandprofile).   
Note that $g_{\tbox{SLRT}}\sim q$  
is the simplest guess that reflects    
the majority dominance expectation.

%%%%%%%%%%%%%%%%%%%%%%%%%%%%%%%%%%%%%%%%%%%%%%%%%%%%%%%%%%%%%%%%
%%%%%%%%%%%%%%%%%%%%%%%%%%%%%%%%%%%%%%%%%%%%%%%%%%%%%%%%%%%%%%%%
%%%%%%%%%%%%%%%%%%%%%%%%%%%%%%%%%%%%%%%%%%%%%%%%%%%%%%%%%%%%%%%%
\section{The semiclassical estimate of the Ohmic conductance}
\label{sec:The semi-classical estimate}

There is a well established semi-classical procedure 
to deduce the algebraic average of the matrix elements $|V_{mn}|^2$
that correspond to the energy difference $\omega=E_m{-}E_n$ 
from the associated correlation function $\langle V(t) V(0) \rangle$.
We would like to apply this procedure in order 
to estimate the conductance of metallic rings. 
Hence our interest is in the matrix elements of 
the velocity operator. The semi-classical estimate  
is based on the following observation:
\be{0}
\langle\langle |v_{mn}|^2 \rangle\rangle_{\omega} 
\ \ = \ \ 
\frac{1}{2\pi\varrho_{\tbox{E}}}
\mbox{FT}\Big[ \langle v(t) v(0) \rangle \Big] 
\ee
where FT stands for Fourier transform.
The velocity-velocity correlation function 
can be obtained via a time-derivative of the time 
dependent diffusion coefficient~$\mathcal{D}(t)$, 
which is the time derivative of the 
spreading~${\langle(r(t)-r(0))^2\rangle}$.

In the Drude ``classical'' approximation 
one assumes an exponential decay of the 
velocity-velocity correlation function, 
and long time diffusion $\mathcal{D}_0$  
as determined by the mean free path 
(see App.~\ref{app:Drude}).
This is satisfactory in the ballistic and diffusive 
regimes, and leads to a Lorentzian line shape:
\be{101}
\langle\langle |v_{mn}|^2 \rangle\rangle_{\omega}  
= \frac{1}{b} v_{\tbox{E}}^2
\frac{R(\omega)}{1+(t_{\ell}\omega)^2}
\ee
where $b=\mathcal{M}L/\ell$ is the dimensionless 
bandwidth of the matrix, and $R(\omega)=1$.

But  in the Anderson regime we know that there 
is a breaktime $t_{\tbox{loc}}$ that marks the 
crossover form diffusion to localization, 
and hence for a bulk system formally ${\mathcal{D}=0}$. 
Consequently, the FT consideration of Eq.~(\ref{e101})
leads to the conclusion that in the limit ${\omega\rightarrow0}$ 
the band-profile should vanish if the system is infinite. 
The simplest reasoning \cite{kbv} leads to the expression 
\be{102}
R(\omega) \ \ = \ \ \frac{L}{\ell_{\xi}}
\eexp{-2L/\ell_{\xi}} \ + \
\frac{1}{1+(t_{\tbox{loc}}\omega)^{-2}}
\ee
where the first term reflects the finite length  
of the system and it is deduced using Eq.~(\ref{e74}). 
We shall see in Sec.\ref{sec:RMT Anderson}, 
using a different more refined approach,  
that the second term is almost correct. 
Namely, the more careful analysis using the Mott's 
picture predicts that the small frequency 
dependence is not $[\omega]^2$ but $[\omega\log(\omega)]^2$.

The quantum mechanical analysis 
should further take into account 
{\bf (i)} the statistics of the levels 
and {\bf (ii)} the fluctuations in 
the size of the matrix elements. 
The former implies wiggles in $R(\omega)$ for small frequencies, 
while the latter imply that the average  
size of the matrix elements does not necessarily 
reflect their typical value.  
Fig.~\ref{fig:time_scales} summarizes the dependance 
of the matrix elements on the disorder.

It should be clear that we always 
have the sum rule 
\be{103}
\sum_{m}|v_{mn}|^2 \ \ = \ \ v_{\tbox{E}}^2
\ee
In the clean ring limit the sum is dominated 
by the diagonal or near diagonal element, 
while all the other off-diagonal elements become 
negligible. Still the estimate Eq.~(\ref{e101}) 
for the other off-diagonal matrix elements 
remains valid and can be justified using 1st order 
perturbation theory.
If the ring is ballistic (but not "clean") 
then the semi-classical estimate Eq.~(\ref{e101})  
implies that the large elements 
form a band of width ${b>1}$.  
If the matrix is not sparse, 
then the contribution of all the $b$ 
in-band elements to the sum rule is comparable. 
But if (say) only a fraction~${s\ll 1}$ 
of of elements are contributing, then their typical value
${|v_{mn}|^2 \sim \langle\langle |v_{mn}|^2 \rangle\rangle_{\omega}/s}$ 
is much larger compared with the average.  
We shall come back to a more detailed 
discussion of `sparsity' in the subsequent sections.

In the diffusion regime $R(\omega)$ mainly reflects 
the level spacing statistics of the individual levels, 
which is a ``microscopic" effect 
that leads to small weak localization corrections 
that had been studied extensively \cite{G1,kamenev}.  
But in the strong localization Anderson regime   
the implication of the breaktime leads 
to the dramatic conclusion that ${R(\omega)\ll1}$  
for ${\omega \ll \Delta_{\xi}}$, 
where the local level spacing ${\Delta_{\xi}}$ 
is {\em not} related to the volume dependent 
microscopic level spacing $\varrho_{\tbox{E}}^{-1}$, 
but to the strength of the disorder.

In the Anderson regime it is evident that 
${\langle\langle |v_{mn}|^2 \rangle\rangle_{\omega}}$
is not the typical value of the matrix elements. 
Roughly speaking, and disregarding the $\omega$ dependence,   
\be{95}
|v_{mn}| \ \ \sim \ \ \frac{v_{\tbox{E}}}{\mathcal{M}} 
\,\exp\left(-\frac{|r|}{\ell_{\xi}}\right)
\ee
where $r\in[0,L/2]$ has a uniform distribution, 
implying a log-box distribution for the size of the elements. 
Accordingly, the typical value is exponentially 
small in the length of the ring, while the average 
is determined by the small percentage of large elements, 
and comes out in agreement with the semi-classical estimate. 
In Sec.~\ref{sec:RMT Anderson} we further elaborate on the statistical 
analysis of the sparsity in the Anderson regime 
using the Mott's picture of localization.

%%%%%%%%%%%%%%%%%%%%%%%%%%%%%%%%%%%%%%%%%%%%%%%%%%%%%%%%%%%5
\section{The RMT statistics in the ballistic regime}
\label{sec:RMT ballistic}

\begin{figure}[h!]

\includegraphics[clip, width=0.7\hsize]{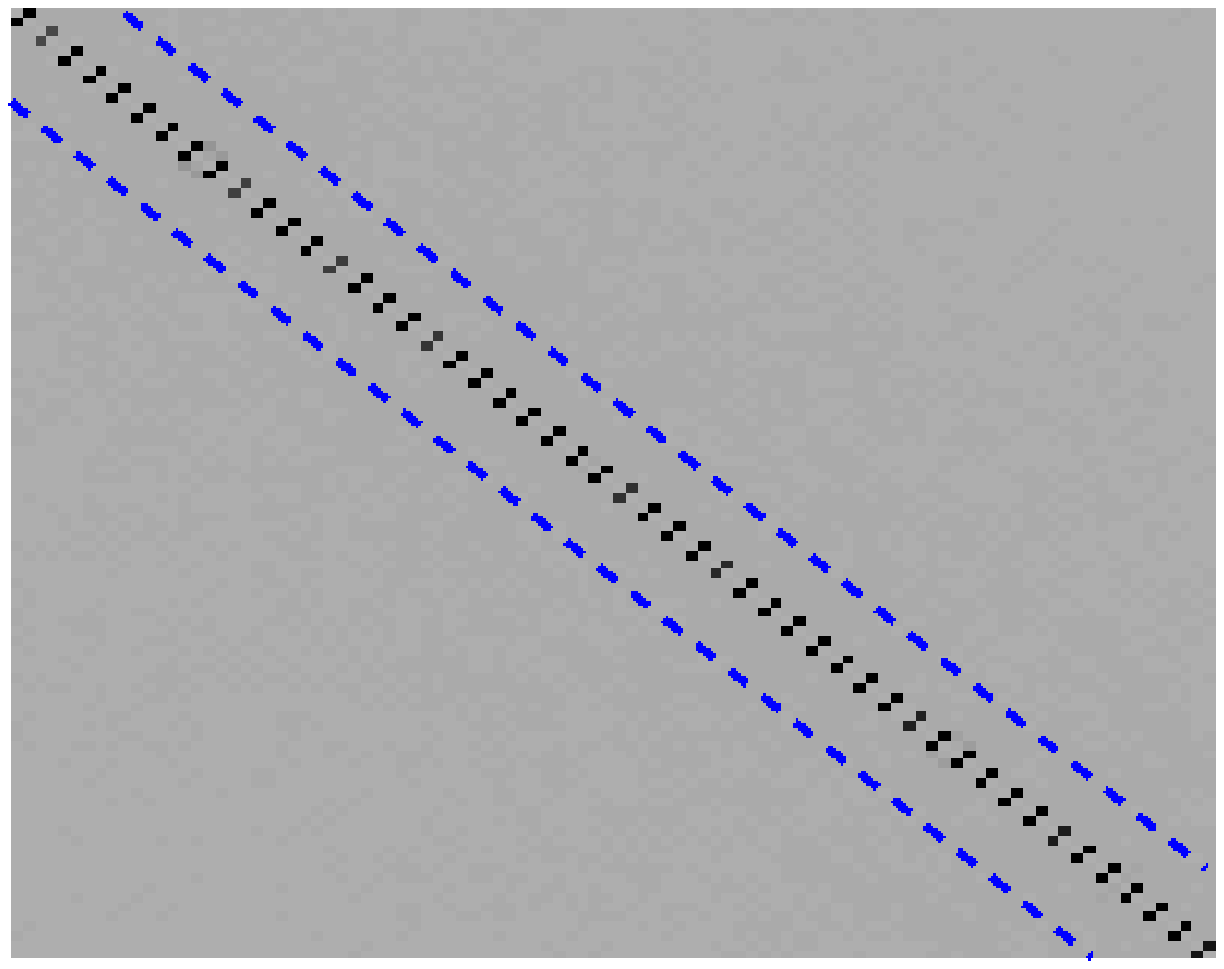} \\
\vspace*{1mm}
\includegraphics[clip, width=0.7\hsize]{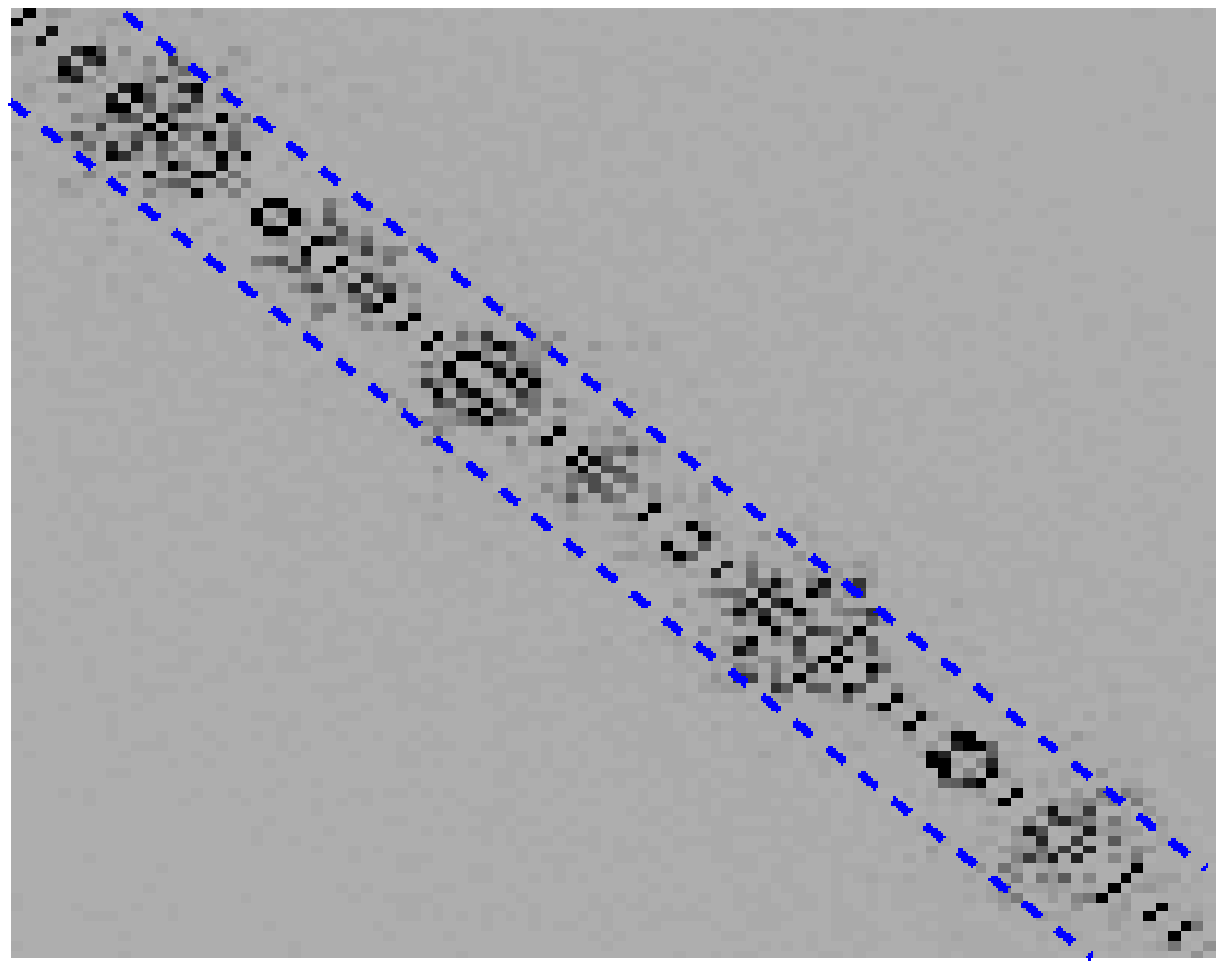}

\caption{
Images of $100\times 100$ pieces of the perturbation matrix $|v_{nm}|^2$ 
in the clean ($W=0.001$)  and ballistic ($W=0.1$) regimes.
The dashed lines correspond to the ballistic bandwidth $\mathcal{M}=10$,   
which is associated with the time scale $t_L$.  
}

\label{fig:mat_ballistic}
\end{figure}

For zero disorder ${W=0}$ each energy level is doubly
degenerate in the basis of real eigenfunctions, 
and the couplings are pairwise, i.e. the matrix element 
between states of different energies is zero.  
See Fig.~\ref{fig:mat_ballistic}. Consequently, the EMF 
cannot induce connected sequences of transitions, 
and the SLRT conductance should be zero. 
The non-zero elements of the perturbation matrix according 
to the sum rule (Eq.~\ref{e103})
are ${|v_{nm}| = v_{\tbox{E}}}$.
The algebraic average of the near diagonal elements 
equals this value (of the large size elements) 
multiplied by their percentage~$p_0 \approx 1/2$. 
Consequently,  
\be{0}
\langle\langle |v_{nm}|^2 \rangle\rangle_{\tbox{a}} 
\approx 
\frac{1}{2} v_{\tbox{F}}^2
\ee
For sufficiently small~$W$ these large size matrix elements
are not affected, and therefore, the algebraic 
average stays the same. Consequently, in the clean ring limit 
the Kubo conductance is formally finite, and attains 
the maximal value as discussed with regard to Eq.~(\ref{e85}).

\begin{figure}[h!]

\includegraphics[clip, width=0.9\hsize]{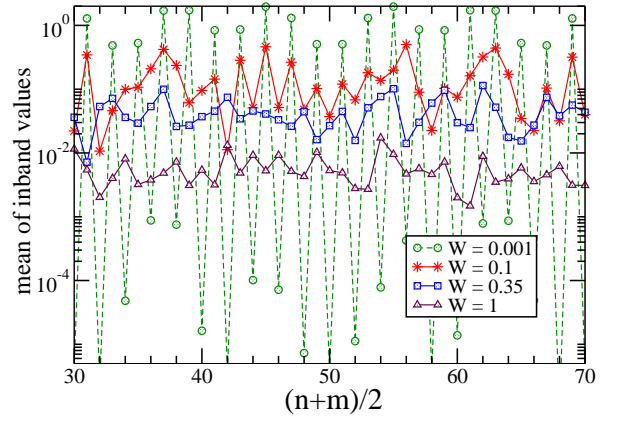}

\caption{
The mean of the in-band values of the $|v_{nm}|^2$ matrix elements 
as a function of~$(n{+}m)/2$ for different values of disorder. 
The pronounced modulation in the ballistic regime is an indication for texture.
}

\label{fig:Vnm_elevel}
\end{figure}

\begin{figure}[h!]

\includegraphics[clip, width=0.9\hsize]{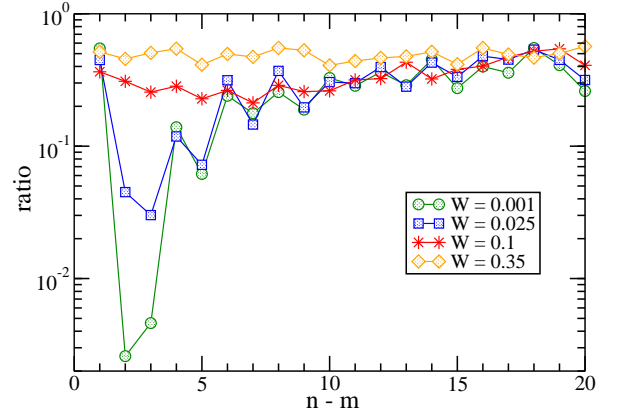}

\caption{
The ratio between the typical value  $[|v_{nm}|^2]_{\omega}$
and the average value $\langle |v_{nm}|^2\rangle_{\omega}$ 
that enter into Eq.~(\ref{e728}) and Eq.~(\ref{e536}), respectively.
In this plot the typical value is the median. 
The horizontal axis is $n{-}m$ corresponding 
to the scaled frequency $\varrho_{\tbox{E}}\omega$. 
Note that both $[|v_{nm}|^2]_{\omega}$ and $\langle |v_{nm}|^2\rangle_{\omega}$ 
when plotted as a function of $\omega$ have a Lorentzian line shape. 
}

\label{fig:func_vs_omega}
\end{figure}

In the clean as well as in the whole ballistic regime 
the algebraic average $\langle\langle |v_{mn}|^2 \rangle\rangle_{\omega}$
does not reflect the sparsity and the textures of the $v_{mn}$.
See Figs.~\ref{fig:Vnm_elevel} and~\ref{fig:func_vs_omega}.
When we look on the image of $v_{mn}$ the immediate 
reaction is to be impressed by the {\em texture}, 
and therefore we discuss it first. Subsequently, we 
discuss the {\em sparsity}, which is in fact more significant 
for the analysis.

The mean DOS of the 2D ring is $\varrho_{\tbox{E}}$.
But $L_x\gg L_y$ and, therefore, it is not uniform.
As the disorder $W$ is increased, 
levels start to mix first in the high DOS regions, 
and only later in the low DOS regions. 
This is the reason for the appearance of textures.  
Let us be more detailed about the non-uniformity 
of the DOS. As a function of the energy~$E$ each time 
that a mode is opened the DOS is boosted.
Consequently, $\varrho_{\tbox{E}}$ is modulated.
This systematic modulation is associated 
with the opening of a single additional mode
at every threshold energy and, therefore, scales 
like $1/\mathcal{M}$. On top there is 
an additional weaker non-systematic modulation 
of the DOS, because the levels of low density modes 
add up to the levels of the high density modes. 
It is the latter type of modulation which is reflected  
in Figs.~\ref{fig:mat_ballistic} and \ref{fig:Vnm_elevel}, 
where the energy window contains throughout exactly $10$ open modes.

In the regions where levels are not yet mixed 
one can estimate the majority of small matrix elements 
using first order perturbation theory: 
Due to the first-order mixing 
of the levels, the typical overlap  ${|\langle \bm{m} | n \rangle|}$  
between perturbed and unperturbed states is
\be{0}
|\langle \bm{m} | n \rangle| 
= \left|\frac{U_{\bm{n}\bm{m}}}{E_{\bm{n}} - E_{\bm{m}}}\right|
\ee
The typical size of a small $v_{nm}$ element 
is the multiplication of this overlap, 
calculated for ${(E_{\bm{n}}{-}E_{\bm{m}}) \sim \varrho{\tbox{E}}^{-1}}$, 
by the size of the non-zero  ${|v_{\bm{n}\bm{m}}|=v_{\tbox{E}}}$ 
element. As a-priori expected this first order estimate 
gives a result that agrees with the semi-classical estimate Eq.~(\ref{e101})
evaluated for $\omega\sim \varrho_{\tbox{E}}^{-1}$. Thus 
\be{111}
q \ \ \approx \ \ \mathcal{M} \frac{L}{\ell}
\ \ \ \ \ \ \ \ \ \ \mbox{[for white disorder]}
\ee
Above some threshold, first order perturbation theory
fails everywhere, meaning that non-perturbative mixing
takes place in any energy. Still, due to the modulation
of the DOS, the mixing range is wider in the near-thresholds
energies, and therefore the matrix elements there
are {\em smaller}. So now we have the opposite situation,
of {\em high~DOS~bottleneck} instead of {\em low~DOS~bottleneck}.

One easily observes that the crossover from weak disorder
(that features separated mixing regions and low~DOS~bottlenecks)
to stronger disorder (that features a connected mixing region
and high~DOS~bottlenecks) is associated with the crossover
from the "clean" to the "ballistic" regime.
The width of the crossover region depends on the
non-uniformity of the DOS, and therefore diminishes 
as the number of open modes becomes large.

%%%%%

The above reasoning implies that the texture might be
important in the SLRT analysis primarily in the clean
ring regime, but much less in the genuine ballistic regime.
But what about sparsity?
Using the FGR in order to determine 
the energy range over which mixing
takes place, we obtain an estimate 
for the bandwidth of the perturbation matrix
\be{0}
b \ = \ 2\pi\varrho_{\tbox{E}}^2 |U_{nm}|^2 
\ \approx \ \mathcal{M}\frac{L}{\ell}
\ee
which agrees with the semi-classical estimate.
But in the ballistic regime ${b<\mathcal{M}}$.
This means that a typical eigenstates cannot occupy all
the $\mathcal{M}$ open modes. Rather it has there
a participation number~${M=b}$ smaller than~$\mathcal{M}$ 
(see Fig.~\ref{fig:measures}).
Consequently, we deduce that the sparsity of
the perturbation matrix is 
\be{113}
s \ = \ \frac{M}{\mathcal{M}} \ \approx \ \frac{L}{\ell} 
\ \ \ \equiv q^2
\ \ \ \ \ \ \ \mbox{[for smooth disorder]}
\ee
where the identification of $s$ with $q^2$ is 
based on the assumption of a log-normal distribution 
which we further discuss in the next paragraph. 
Unlike the texture, the sparsity persists
via the whole ballistic regime up to the border
with the diffusive regime. For this reason 
we regard the sparsity as the main ingredient 
in the SLRT analysis.

The discussion of sparsity in the previous paragraph 
is somewhat meaningless unless one specifies the  
distribution to which $s$ refers.
At this point of the discussion it is essential to distinguish 
between {\em white disorder} for which the scattering 
is isotropic, and {\em smooth disorder} for which 
only nearby modes are coupled (small scattering angle). 
The latter applies if the potential floor within the ring 
has a smooth rather than erratic variation with respect to 
the Fermi wavelength.
Assuming smooth disorder it becomes essential to 
extend the perturbation theory of App.~\ref{app:model} 
beyond first order. It makes sense to say 
that $|v_{mn}|\sim |W|^r$,
where the order~$r$ is bounded by~$\mathcal{M}$.
Therefore, $\log(|v_{mn}|)$ has some bounded
distribution which can be approximated (say) by a Gaussian.
It follows that a log-normal ensemble should be
qualitatively appropriate to describe the statistical properties.
It follows that $g_{\tbox{SLRT}}$ can be estimated 
using Eq.~(\ref{e65}) with the $q$ of Eq.~(\ref{e113}).
On the other hand in the case of white disorder $|v_{mn}|$ 
of the majority elements is given by {\em first} 
order perturbation theory, and then one should 
use Eq.~(\ref{e65}) with the $q$ of Eq.~(\ref{e111}).

%%%%%%%%%%%%%%%%%%%%%%%%%%%%%%%%%%%%%%%%%%%%%%%%%%%%%%%%%
\section{The RMT statistics in the Anderson regime}
\label{sec:RMT Anderson}

\begin{figure}[h!]

\includegraphics[clip, width=0.9\hsize]{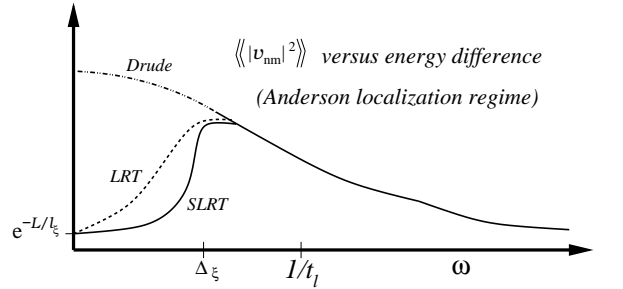}

\caption{
Schematic plot that illustrates the dependence 
of the average and the typical values of the matrix elements
on the energy separation~$\omega$.  
These are labeled as `LRT' and `SLRT' respectively 
and compared with the semiclassical (`Drude') expectation.
The plot refers to the Anderson regime.  
For a corresponding numerical illustration 
in the ballistic regime see Fig.\ref{fig:func_vs_omega}. 
}

\label{fig:disorder_regimes}
\end{figure}

The simplest picture of localization regards
the lattice as composed 
of segments of size $\ell_{\xi}$, 
and assumes that each eigenstate is   
well localized in one of this segments. 
Accordingly, non-negligible matrix elements 
are only between states that reside 
in the same space segment. We shall refer 
to this as the ``zero order'' picture.
Taking into account that 
the matrix element of the velocity 
operator are related to those of the position operator 
by the relation $|v_{mn}|^2=\omega^2|r_{mn}|^2$, 
it follows that $R(\omega) \sim (\omega/\Delta_{\xi})^2$
in consistency with the semi-classical reasoning  
of Sec.~\ref{sec:The semi-classical estimate}, 
which is summarized by Fig.~\ref{fig:disorder_regimes}.

In order to refine this picture we use the 
following procedure due to Mott. 
The zero order basis is determined 
by ignoring the possibility of the particle 
to hop from segment to segment. 
In order to find the "true" eigenstates  
we have to take into account the  
residual interaction. It is reasonable  
to postulate that if the distance between 
two zero order eigenstates is ${r=r_n-r_m}$, 
then the residual interaction is 
\be{0}
\kappa = \Delta_{\xi} \exp(-|r|/\ell_{\xi})
\ee 
The prefactor is the natural educated guess,  
which is later justified (see below) by requiring 
consistency with the semi-classical result.

If we have two zero-order eigenstates 
that do not reside at the same segment, 
but have distance $r$ in space  
and distance $\varepsilon$ in energy, 
then the true eigenstates have energy 
difference $\omega = \sqrt{\varepsilon^2+\kappa^2}$, 
and the dipole matrix element becomes 
${|r_{mn}| = (\kappa/\omega) \times (r/2)}$ instead of zero.
Originally the zero-order eigenstates 
had a density ${(\varrho_{\tbox{E}}/L) d\varepsilon dr}$,  
but now the region $|\varepsilon|<\exp(-|r|/\ell_{\xi})$ 
is depleted, and forms a density $d\omega/\Delta_{\xi}$ 
of so called Mott resonant states. 
If we slice all those states that have energy 
difference $\omega$ then  
\be{0}
|v_{mn}| 
\ \ \sim \ \ 
\left\{\amatrix{
\Delta_{\xi} \ r \ \eexp{-|r|/\ell_{\xi}} 
& \ \ \ \ \ \ \mbox{off res.} \cr
|\omega| \ r_{\omega}  
& \ \ \ \ \ \ \mbox{on res.} 
}\right. 
\ee
where 
\be{0}
r_{\omega} = \ell_{\xi}\log(\Delta_{\xi}/\omega)
\ee
This implies that the size distribution of the $|v_{mn}|$ 
elements that reside inside a band of width $\omega$ is within 
\be{0}
\frac{v_{\tbox{E}}}{\mathcal{M}}\times
\Big[
\frac{L}{\ell_{\xi}}\eexp{-L/\ell_{\xi}}, 
&  1
\Big] 
& \ \ \ \ \ \ \mbox{for $|\omega|>\Delta_{\xi}$} \ \ \ \  
\\
\frac{v_{\tbox{E}}}{\mathcal{M}}\times
\Big[
\frac{L}{\ell_{\xi}}\eexp{-L/\ell_{\xi}}, 
& \frac{\omega}{\Delta_{\xi}} 
\log\left(\frac{\Delta_{\xi}}{\omega}\right)
\Big]
& \ \ \ \ \ \ \mbox{for $|\omega|<\Delta_{\xi}$} \ \ \ \ 
\ee 
Compared with Eq.~(\ref{e102}) this is a refinement 
that takes properly into account the $\omega$ dependence 
of the matrix elements. Disregarding a logarithmic 
correction it reproduces the semi-classical result Eq.~(\ref{e101}).    

If we ignore the Mott resonant states, 
then a {\em log-box} distribution is 
implied. The Mott resonant states form 
a {\em box} distribution on top. 
In a log-scale the Mott resonant states 
contributes a peak of large elements. 
But this peak does not affect the $x_{\omega}$
calculation. Consequently, for practical purpose 
we can regard the matrix elements in the SLRT calculation 
as having a simple log-box distribution 
as reflected by the crude approximation of Eq.~(\ref{e95}). 
The sparsity of this distribution is characterized by  
\be{633}
\tilde p \ \ = \ \ \mathcal{M}\frac{\ell}{L}
\ee
and the SLRT suppression factor is given by Eq.~(\ref{e63}).

%%%%%%%%%%%%%%%%%%%%%%%%%%%%%%%%%%%%%%%%%%%%%%%%%%%%%%%%
\section{SLRT vs VRH calculation}
\label{sec:SLRT vs VRH}

In order to appreciate the similarities  
and the differences between SLRT  
and the conventional Hopping calculation, 
we cast the latter into the SLRT language. 
Eq.~(4.4) of Ref.\cite{AHL} for the DC Hopping conductance 
due to phonon induced transitions is 
\be{0}   
G_{\tbox{Ohm}} \ \ = \ \ 
\frac{1}{\mathcal{N}}
\left[\left[
\frac{e^2}{T}(1{-}f(E_n))f(E_m)w^{\gamma}_{mn}
\right]\right]_{\parallel} 
\ee   
The notation $[[...]]_{\parallel}$ implies that 
the resistance of the network is calculated 
between states at the same energy ${E \sim E_{\tbox{E}}}$,  
that reside in opposite sides of the sample.
Due to the Fermi occupation factor, the network 
contains effectively $\mathcal{N}=\varrho_{\tbox{E}}T$ nodes.  
The division by $\mathcal{N}$ is required because 
we have defined the $[[...]]$ as inverse-resistivity 
and not as inverse-resistance of the network.

The occupation factor $(1{-}f(E_n))f(E_m)/T$  
gives $\mathcal{O}(1)$ weight only to the $\mathcal{N}$ levels  
that reside within a window of width $T$. 
If we ignore the relaxation effects 
and regard the fluctuating environment 
as a noise source that induces transitions 
${w^{\gamma}_{mn}\propto \exp(|E_m{-}E_n|/T)}$, 
we still should get the same result for $G$,  
even if we omit the occupation factor.
This point of view allows to bridge between 
the noisy driving problem that we consider 
in this paper and the phonon-induced hopping 
in the prevailing literature.

The Einstein relation $G_{\tbox{Ohm}}=(e/L)^2\varrho_{\tbox{E}}\mathcal{D}$ 
relates the conductance and diffusion in real space. 
We deduce that 
\be{0}   
\mathcal{D} \ \ = \ \ 
\left(\frac{L}{\mathcal{N}}\right)^2
\left[\left[
w^{\gamma}_{mn} 
\right]\right]_{\parallel} 
\ee   
This should be compared with the SLRT expression 
for the noise induced energy diffusion 
\be{0}   
\mathcal{D}_{\tbox{E}} \ \ = \ \ 
\left(\frac{1}{\varrho_{\tbox{E}}}\right)^2
\left[\left[
w^{\gamma}_{mn} 
\right]\right]_{\perp} 
\ee   
Here the resistance of the network is 
calculated between states that reside 
far away in energy. 
The SLRT result for $\mathcal{D}_{\tbox{E}}$ 
and the hopping implied result for $\mathcal{D}$ 
are both simple and manifestly equivalent: 
The diffusion coefficient equals the transition 
rate $[[w^{\gamma}_{mn}]]$ times the {\em step} squared. 
In the SLRT calculation the step in 
energy space is $1/\varrho_{\tbox{E}}$, 
while in the standard real space analysis  
the step is $L/\mathcal{N}$. 
Optimization of the hopping with respect 
to the distance~$\omega$ in energy  
is equivalent to optimization
with respect to the distance~$r$ in space.

%%%%%%%%%%%%%%%%%%%%%%%
\section{Can we get VRH from Kubo?}
\label{sec:VRH from Kubo?}

The analysis that we have introduced in this paper gives the impression
that SLRT is essential in order to derive the VRH result.
This statement looks to be in contradiction with 
the prevailing common wisdom, and therefore deserves 
further clarification. In the discussion below we explain 
that VRH can be obtained from Kubo for an {\em artificial} 
toy model, but not for the {\em physical} model that 
we have analyzed in this paper following Anderson and Mott.

It is instructive to point out that the Kubo formula Eq.~(\ref{e24})
can be rephrased as saying that 
\be{0}
D \ \ = \ \ [D_n]_{\tbox{avr}}
\ee
where $D_n$ is the diffusion coefficient for 
a spreading process that start at state~$n$. 
If we consider an artificial model where the eigenstates 
are labeled as ${n=(i\nu)}$, with energies ${E_{i\nu}=\epsilon_{\nu}}$ 
and matrix elements ${V_{i\nu,j\mu}\sim\exp(-|r_i-r_j|/\ell_{\xi})}$,
such that
\be{0}
w_{i\nu,j\mu} 
\ \ \sim \ \ 
\exp\left[ 
-\frac{|r_i-r_j|}{\ell_{\xi}} 
-\frac{|\epsilon_{\nu}-\epsilon_{\mu}|}{\omega_c}    
\right]
\ee
then all the $D_n$ are the same value. 
Furthermore, their common value is given by a VRH-like expression
which reflects an optimization of the hopping distance.
Consequently, the average $D$ is also given 
by the exactly the same VRH-like expression.

However, in the physical model that we have considered 
in this paper the $D_n$ in the Anderson regime 
are typically dominated by one term only, 
and therefore wildly fluctuate. 
It is then clear that an algebraic average 
would give a very large result which is dominated 
by the minority of large elements. 
In fact our analysis, which merely reproduces Mott's original analysis, 
shows that up to logarithmic correction the Kubo formula 
gives $G\propto\omega_c^2$.  In order to get VRH we have 
to perform an SLRT analysis rather than LRT analysis.

In the above discussion one could wonder whether a good 
strategy for obtaining an SLRT estimate would be to take 
a harmonic instead of algebraic average over $D_n$. 
In fact there are circumstances where such procedure 
gives a very good result \cite{kbr}.
However, in general such procedure is expected 
to underestimate the correct result, because it is based on the 
assumption that the hopping is always with the same 
optimal step, as in series addition of resistors,  
without the possibility to bypass in parallel.

%%%%%%%%%%%%%%%%%%%%%%%
\section{VRH vs Hopping}
\label{sec:VRH vs Hopping}

It is customary to assume that a noisy non-thermal 
source has a Lorentzian power spectrum:
\be{0}
\tilde{F}(\omega) = \frac{1}{\pi}
\frac{\omega_c}{\omega^2+\omega_c^2}
\ee
Let us consider the Anderson regime and assume 
that $\omega_c \ll \Delta_{\xi}$. 
It should be clear that the VRH result is not 
applicable here. This is because the transport 
is dominated by ${\omega>\Delta_{\xi}}$ 
transitions. In this case SLRT give the same 
result as Kubo, which we call simple hopping \cite{kbv}:
\be{19}
\mathcal{D} \ \ \approx \ \ 
\omega_c t_{\tbox{loc}} \, \mathcal{D}_0
\ \ = \ \ \frac{(\ell_{\xi})^2}{t_c}  
\ee
where ${t_c=1/\omega_c}$.
This is as expected from heuristic considerations. 
It describes a random walk hopping process with 
steps of size $\ell_{\xi}$ and time $\tau_{\gamma}$.  
This type of result has been highlighted 
in old studies of the quantum kicked rotator problem \cite{qkr}.

%%%%%%%%%%%%%%%%%%%%%%%
\section{Experimental demonstration of semi linear response}
\label{sec:experiment}

For a given metallic ring the experimentalist has control 
over the frequency and on the strength of the driving. 
These can be adjusted such that FGR transitions are the dominant 
mechanism for energy absorption. This excludes the adiabatic 
regime where near neighbor transitions dominate 
either due to Landau-Zener \cite{wilk} or Debye relaxation mechanism \cite{rings}.   

Assuming that FGR transitions are the dominant 
mechanism, this does not automatically imply 
linear response. The rate of the driven transitions 
can be smaller or larger compared with the environmental 
induced rate of transitions, and accordingly we expect 
a crossover from LRT to SLRT \cite{kbr}.

The simple minded indication for semi-linear response  
is a drop in the value of the absorption coefficient 
if the driving is strong enough (see estimates below).  
What can be measured is the SLRT suppression 
factor $g_{\tbox{SLRT}}$ and its dependence on 
the spectral content of the driving.

As observed in Ref.\cite{slr}, the distinction 
of semi-linear from linear response 
is not ambiguous. The theory is called SLRT because on the one hand  
the power spectrum ${\tilde{S}(\omega) \mapsto \lambda \tilde{S}(\omega)}$ 
leads to ${D \mapsto \lambda D}$, but on the other 
hand ${\tilde{S}(\omega) \mapsto \tilde{S}_1(\omega)+\tilde{S}_2(\omega)}$ 
does not lead to ${D \mapsto D_1 + D_2}$. 
This semi-linearity can be tested in an experiment 
in order to distinguish it from linear response.

Let us discuss in more details the experimental 
conditions that are required in order to observe 
semi-linear response. The problem is characterized 
by the following parameters:
\be{0}
\mbox{system} &:& (\omega_0, \omega_{c}^{\tbox{sys}}) \\
\mbox{driving} &:& (\omega_{c}, \varepsilon) \\
\mbox{bath} &:& (\gamma_{\phi}, \gamma_{\tbox{rlx}})
\ee
where $\omega_{c}^{\tbox{sys}}$ is the frequency 
that characterizes the semi-classical motion;  
$\omega_0=\varrho_{\tbox{E}}^{-1}$ is the frequency 
corresponding to the mean level spacing; 
$\omega_{c}$ and $\varepsilon$ are the cut-off frequency
and the RMS value of the driving (EMF); 
and $\gamma_{\phi}, \gamma_{\tbox{rlx}}$ 
are the dephasing and the relaxation rates due to the environment.

As already stated we are not interested 
in adiabatic driving ($\omega_c < \omega_0$)
but rather in what we call DC driving. 
The conditions that have to be satisfied in 
an SLRT oriented experiment are:
\be{0}
%
%\mbox{Mesoscopic regime} &:& \omega_0 \ll \gamma \ll 
%\omega_{c}^{\tbox{sys}}\\
%
\mbox{DC driving} &:& \omega_0 \ll \omega_c \ll
\omega_{c}^{\tbox{sys}}\\
\mbox{FGR condition} &:& \omega_0 \ll w_{\varepsilon} \ll \omega_c \\
\mbox{LRT condition} &:&  w_{\varepsilon} \ll \gamma \\
\mbox{SLRT condition} &:& \gamma \ll w_{\varepsilon} 
\ee
where $w_{\varepsilon}$ is the FGR transition rate (Eq.~(\ref{e41})).

There are several experimental methods which could support the theoretical
predictions of our paper. The experiment can be based on metallic rings
(gold \cite{webb,Bluhm}, copper \cite{Levy}, silver \cite{Mailly_silver}), GaAs
and other semiconductor heterostructures (\cite{Mailly, Lee}), molecular wires,
etc. To estimate the experimental numbers let us consider a semiconductor (GaAs)
ring driven by time-dependent magnetic flux
\be{0}
&{\mathcal{M} = 5}, \ \ \
{L = 0.1 \ \mu m}, \ \ \ 
{\ell = 50 \ \mu m} \\
&v_{\tbox{F}} = 2.7\times 10^5\ m/s
\ee
The long mean free path is required in order to be 
deep in the ballistic regime with sparsity  
\be{0}
q \ = \ \mathcal{M}\frac{L}{\ell} \ \sim \ 0.01
\ee
By Eq.~(\ref{e23}) the mean level spacing is 
\be{0}
\omega_0 \ = \ \frac{2 v_{\tbox{F}}}{\mathcal{M} L} \ \approx \ 1 \ m eV
\ee
The ballistic time is ${t_{L} = L/v_{\tbox{F}} \approx 3.7\times10^{-13}\ s}$ hence
\be{0}
{\omega_{c}^{\tbox{sys}} = \frac{2\pi v_{\tbox{F}}}{L} \approx 11\ meV}
\ee
which is $\sim 10^{13}\ Hz$ in frequency units.
%
%
% In order to be in the Mesoscopic regime 
% we assume a relaxation time ${t_{\tbox{rlx}} \sim 2\times 10^{-12}\ s}$ 
% corresponding to  ${\gamma \sim 3\omega_0 } $.
%
In order to satisfy the DC driving condition we assume 
a power spectrum of width $\omega_c \lesssim \omega_{c}^{\tbox{sys}}$. 
The EMF is induced by a time-dependent magnetic field $\varepsilon \approx (\omega_c L^2)B$.
The FGR rate is estimated using Eq.~(\ref{e41}) with $E_n{-}E_m \sim \omega_0$
\be{0}
w_{\varepsilon} 
\ \approx \ 
\frac{e^2 \mathcal{M}^3 L}{\omega_c \ell} \times \varepsilon^2
\ \approx \ \frac{e^2 \mathcal{M}^3 L^5 \omega_c}{\ell} \times B^2
\ee
In order to satisfy the FGR condition the
magnetic field should be at least $180\ G$.
The expected crossover between linear to semi-linear 
response occurs for $w_\varepsilon \sim \gamma$.
Assuming for example ${t_{\tbox{rlx}} \sim 2\times 10^{-12}\ s}$ 
we get ${\gamma \sim 3\omega_0}$ leading to 
\be{0}
B_{\tbox{SLRT threshold}}  \ \ \sim \ \ 320\ G
\ee
Under the above conditions we expect that as $B$ is increased 
there will be crossover from linear to semi-linear response   
with suppression factor ${g_{\tbox{SLRT}} \approx 0.3}$, 
where we used Eq.~(\ref{e65}).  The crossover is of course 
not sharp because $w_{\varepsilon}$ is after all distributed over 
a wide range. In fact the functional shape of the crossover 
can be used in order to deduce this distribution \cite{kbb}. 
In any case it should be re-emphasized that the experimental verification 
for the nature of the crossover requires merely to test whether the absorption rate 
depends in a non-linear way on the spectral content of the driving.

%%%%%%%%%%%%%%%%%%%%%%%
\section{Summary and Discussion}
\label{sec:Summary}

Possibly the nicest thing about SLRT is that it consists 
a natural extension of LRT, that places under one roof 
various results for the conductance in different regimes. 
It should be clear that in the strict DC limit ($\omega_c\rightarrow0$), 
irrespective of the functional form of the power spectrum, 
we always get for $G$ a result that formally agrees with 
the Landauer formula. See the discussion in Sec.~\ref{sec:The Kubo-Drude
conductance}.
In the diffusive regime it becomes equivalent to the Drude 
formula with small weak localization corrections.  
But in the other regimes (Anderson, Ballistic), if the low 
frequency driving has some arbitrary spectral content, 
then very different results are obtained (Hopping, VRH, generalized VRH).

It is interesting that in our ``minimal'' treatment of the problem 
there is no need to introduce relaxation due to phonons 
in order to get a VRH result. Rather, we regard VRH 
as arising from the competition between the statistical 
properties of the matrix elements and the power spectrum of a noisy driving field.

The formalism allows to take various limits
involving the size of the system ($L$), 
the driving frequency ($\omega_c$) and its intensity ($\varepsilon$), 
and the rate of the environmentally induced transitions ($\gamma$).   
The order of the limits is very important.
In particular:  if we take the limit $L\rightarrow\infty$ 
followed by $\varepsilon\rightarrow0$, 
keeping $\gamma$ constant, then we get LRT; 
while if we take $L\rightarrow\infty$ 
followed by $\gamma\rightarrow0$, 
keeping $\varepsilon$ constant, then we get SLRT. 
Also note that if we keep $L$ constant and 
take $\omega_c\rightarrow 0$ we get the adiabatic 
limit and not the DC limit of LRT / SLRT.

We have dedicated Sec.~\ref{sec:experiment} to introduce 
actual estimates that are required 
in order to observe SLRT in a real experiment. 
It is important to realize that the experimental 
procedure allows to distinguish in a non-ambiguous 
way between LRT and SLRT by playing with the 
spectral content of the driving source. 
Furthermore, one can test specific predictions 
for the $g_{\tbox{SLRT}}$ suppression factor, 
e.g. Eq.~(\ref{e63}) with (\ref{e633}), 
and Eq.~(\ref{e65}) with (\ref{e111}) or (\ref{e113}).  
We note that the explicit incorporation of 
the environmentally induced transitions 
into the resistor-network calculation,  
and the subsequent analysis of the resulting 
SLRT steady state is straightforward \cite{kbb}.

The SLRT calculation is based on a resistor network 
picture of transitions between energy levels,
for which an RMT framework is very appropriate 
and effective.  In the so called ``quantum chaos'' 
context Wigner (in the nuclear context) 
and later Bohigas (in the mesoscopic context) 
have motivated the interest in Gaussian ensembles, 
but there are circumstances where non-Gaussian ensembles
are appropriate, which lead to novel physics. 
Indeed we have faced in this paper the analysis  
of log-normal and log-box ensembles 
corresponding to the weak and strong disorder limits. 
We have demonstrated that for such ensembles 
a large SLRT suppression effect is expected, 
that could not be anticipated within the LRT framework.

%%%%%%%%%%%%%%%%%%%%%%%%%%%%%%%%%%%%%%%%%%%%%%%%%%%%%%%%%%%%%%%%
\ \\

\noindent
{\bf Acknowledgment:} 
We thank Yigal Meir (BGU) and Yuri Galperin (Oslo) 
for discussions of VRH that have motivated some of the 
sections in this paper. This research has been supported 
by a grant from the USA-Israel Binational Science Foundation (BSF).

%%%%%%%%%%%%%%%%%%%%%%%%%%%%%%%%%%%%%%%%%%%%%%%%%%%%%%%%%%%%%%%%%%%%%%%%%%%%%%%%%%%%
%%%%%%%%%%%%%%%%%%%%%%%%%%%%%%%%%%%%%%%%%%%%%%%%%%%%%%%%%%%%%%%%%%%%%%%%%%%%%%%%%%%%
%%%%%%%%%%%%%%%%%%%%%%%%%%%%%%%%%%%%%%%%%%%%%%%%%%%%%%%%%%%%%%%%%%%%%%%%%%%%%%%%%%%%
%%%%%%%%%%%%%%%%%%%%%%%%%%%%%%%%%%%%%%%%%%%%%%%%%%%%%%%%%%%%%%%%%%%%%%%%%%%%%%%%%%%%

\appendix

%%%%%%%%%%%%%%%%%%%%%%%%%%%%%%%%%%%%%%%%%%%%%%%%%%%%%%%%%%%%%%%%
\section{The resistor network calculation}
\label{app:net}

In this appendix we explain how the inverse 
resistivity $G=[[G_{nm}]]$ of a one-dimensional 
resistor network is calculated. We use the language of 
electrical engineering for this purpose.
In general this relation is semi-liner rather than linear, 
namely  ${[[\lambda G]]=\lambda[[G]]}$, 
but ${[[A+B]] \ne [[A]]+[[B]]}$.
The experimental implications of this observation 
in the SLRT context are discussed in Sec.\ref{sec:experiment}.

There are a few cases where an analytical expression 
is available. 
If only near neighbor nodes are connected, 
allowing ${G_{n,n+1}=g_n}$ to be 
different from each other, 
then ``addition in series" implies 
that the inverse resistivity calculated 
for a chain of length~$N$ is  
\be{0}
G \ \ = \ \ \left[\frac{1}{N}\sum_{n=1}^{N} \frac{1}{g_n}\right]^{-1}
\ee
If $G_{nm}=g_{n-m}$ is a function 
of the distance between the nodes $n$ and $m$ 
then it is a nice exercise to prove 
that ``addition in parallel" implies 
\be{0}
G \ \ = \ \ \sum_{r=1}^{\infty} r^2 g_r
\ee

In general an analytical formula for~$G$
is not available, and we have to apply 
a numerical procedure. For this purpose 
we imagine that each node~$n$ is connected 
to a current source $I_n$. The Kirchhoff    
equations for the voltages are    
\be{0}
\sum_m G_{mn} (V_n-V_m)  \  = \ I_n
\ee
This set of equation can be written in a matrix form:
\be{0}
\bm{G} \bm{V}  \ = \ \bm{I}
\ee
where the so-called discrete Laplacian matrix of the network is defined as 
\be{0}
\bm{G}_{nm} =  \left[\sum_{n'}  G_{n'n}\right]\delta_{n,m} - G_{nm}
\ee
This matrix has an eigenvalue zero which is associated  
with a uniform voltage eigenvector. Therefore, it has 
a pseudo-inverse rather than an inverse, and the Kirchhoff 
equation has a solution if and only if ${\sum_n I_n=0}$.      
In order to find the resistance between nodes ${n_{\tbox{in}}=0}$ 
and ${n_{\tbox{our}}=N}$, 
we set ${I_0=1}$ and ${I_N=-1}$ and ${I_n=0}$ otherwise,  
and solve for $V_0$ and $V_N$. 
The inverse resistivity is ${G=[(V_0-V_N)/N]^{-1}}$.

%%%%%%%%%%%%%%%%%%%%%%%%%%%%%%%%%%%%%%%%%%%%%%%%%%%%
\section{Model details}
\label{app:model}

In the numerical study we consider the Anderson tight binding model, 
where the lattice is of size ${L\times M}$ 
with $M \ll L$, and lattice constant~$a$. 
The longitudinal and the transverse hopping amplitudes per unit time 
are ${c_{\parallel}}$ and  ${c_{\perp}}$, respectively.
The random on-site potential in the Anderson 
tight binding model is given by a box distribution of width 
determined by~$W$.

The numerical calculations of Ref.\cite{bld}  
assume ${c_{\parallel}} = 1$ and ${c_{\perp} = 0.9}$.
Thus in the middle of the band there is a finite 
energy window with exactly ${\mathcal{M}=M}$ open modes. 
Rings of length $L = 500$ with $M = 10$ modes has been considered. 
In our re-processed Fig.~\ref{fig:measures} the default 
cutoff is ${\varrho_{\tbox{E}}\omega_c\approx 7}$ as in Ref.\cite{bld},  
but as the disorder becomes weaker it is adjusted such that 
the DC condition ${\varrho_{\tbox{E}}\omega_c \lesssim b}$ is always satisfied.

For white (uncorrelated) disorder 
the Hamiltonian is given by Eq.~(\ref{e7}) with 
the isotropic scattering term
\be{182}
|U_{\bm{n}\bm{m}}|^2  \ \ \approx \ \ \frac{a}{\mathcal{M} L}W^2
\ee
The eigenstates of the Hamiltonian can be found numerically. 
The degree of ergodicity is characterized by the participation 
number $\mbox{PN}\equiv[\sum\rho^2]^{-1}$,  
which is calculated in various representations: 
in position space $\rho_{r_x,r_y}=|\langle r_x,r_y|\Psi\rangle|^2$, 
in position-mode space $\rho_{r_x,k_y}=|\langle r_x,k_y|\Psi\rangle|^2$,  
and in mode space $\rho_{k_y}=\sum_{r_x}|\langle r_x,k_y|\Psi\rangle|^2$, 
where $k_y=[\pi/(\mathcal{M}{+}1)]\times\mbox{\small integer}$.

%%%%%%%%%%%%%%%%%%%%%%%%%%%%%%%%%%%%%%%%%%%%%%%%%%%%
\section{The Drude formula}
\label{app:Drude}

The velocity-velocity correlation function, assuming 
isotropic scattering, is proportional to the survival 
probability $P(t)=\eexp{-t/t_{\ell}}$. 
Ignoring  a factor that has to do with 
the dimensionality $d=2,3$ of the sample 
the relation is 
\be{0}
\langle v(t) v(0) \rangle 
\approx  v_{\tbox{E}}^2 \ P(t)
=  v_{\tbox{E}}^2 \eexp{-|t|/t_{\ell}} 
\ee
The rate of the scattering can be calculated from the~FGR,  
also know as the Born approximation
\be{0}
\frac{1}{t_{\ell}} = 2\pi \varrho_{\tbox{E}} |U_{mn}|^2   
= \frac{\pi a}{v_{\tbox{E}}}W^2
\ee
where in the last equality we used Eq.~(\ref{e182}).
From here we deduce that the mean free path 
(disregarding prefactors of order unity)
\be{0}
\ell \ = \ v_{\tbox{E}}t_{\ell} \ \approx \
\frac{1}{a}\left(\frac{v_{\tbox{E}}}{W}\right)^2 
\ee
and the diffusion coefficient in real space 
\be{0}
\mathcal{D}_0 
\ = \ \frac{1}{2}\int_{-\infty}^{\infty} \langle v(t) v(0) \rangle    
\ \approx \ v_{\tbox{E}}\ell
\ee
By the Einstein relation we deduce the Drude formula 
\be{0}
G_{\tbox{Ohm}} \ = \ \left(\frac{e}{L}\right)^2 \varrho_{\tbox{E}} \mathcal{D}_0
\ = \ \frac{e^2}{2\pi\hbar} \mathcal{M} \ \frac{\ell}{L}
\ee

A literally equivalent route to derive 
the Drude formula is to semi-classically 
deduce $\langle\langle |v_{mn}|^2 \rangle\rangle$
from the velocity-velocity correlation function 
as in Sec.\ref{sec:The semi-classical estimate}, and then to substitute in
Eq.~(\ref{e69}). 
This has the advantage of allowing easy 
generalizations of the Drude formula in the ballistic 
and in the Anderson regimes. 
In this context it is useful to realize \cite{kbf} 
that in the semi-classical picture  
the integral over the velocity-velocity 
correlation function is related to the 
transmission $g_0$ of the ring (if it were dissected).
This leads to the identification in Eq.~(\ref{e74}).

In the diffusive regime Mott has demonstrated 
that it is optionally possible to obtain 
a direct estimate of the dipole matrix elements, 
using a random-wave picture.   
Namely, it is assumed that in the diffusive regime 
the eigenstates of the Hamiltonian are ergodic in position space, 
and look like random waves with a correlation scale~$\ell$. 
Locally the eigenstates are similar
to free waves. The total volume $L^d$
is divided into domains of size $\ell^d$.
Hence we have $(L/\ell)^d$ such domains.
Given a domain, the condition to have
non-vanishing overlap upon integration 
is ${|\vec{q}_n-\vec{q}_m|\ell < 2\pi}$, 
where $\vec{q}$ is the local wavenumber
within this domain. The probability
that $\vec{q}_n$ would coincide
with $\vec{q}_m$ is ${1/(k_E\ell)^{d{-}1}}$. 
The contributions of the non-zero
overlaps add with random signs hence
\be{0}
|v_{mn}| =
\left[ \frac{1}{(k_E\ell)^{d{-}1}} \times \left(\frac{L}{\ell}\right)^d
\right]^{1/2}
\times (\overline{\Psi^2}\ell^d)v_{\tbox{E}}
\ee
Assuming ergodicity $\overline{\Psi^2} \approx 1/L^d$, 
and we get the same estimate as in the semi-classical procedure.

%%%%%%%%%%%%%%%%%%%%%%%%%%%%%%%
\section{The log-box ensemble}
\label{app:log-box}

The cumulative distribution function that corresponds to Eq.~(\ref{e16}) is
\be{0}
\mbox{Prob}(X < x) &=&   \frac{\ln x - \ln x_0}{\ln\left(x_1/x_0\right)}
\ee
The algebraic, geometric and harmonic averages are
\be{0}
\langle\langle x\rangle\rangle_a  &=& \frac{x_1-x_0}{\ln\left(x_1/x_0\right)} \\
\langle\langle x\rangle\rangle_g &=& \sqrt{x_1 x_0} \\
\langle\langle x\rangle\rangle_h &=& \ln\left(x_1/x_0\right) \frac{x_1
x_0}{x_1-x_0}
\ee
Note that for this distribution the median equals the geometric average.
The sparsity parameters are 
\be{0}
s &=& 2\tilde p \ \frac{\eexp{-1/\tilde p}-1}{\eexp{-1/\tilde p}+1}
\\
p &=& -\tilde p \left(\ln \tilde p + \ln \left(1-\eexp{-1/\tilde p} \right) \right) 
\\
q &=& \left(2 \tilde p \sinh \frac{1}{2\tilde p}\right)^{-1} 
\ee
where we defined $\tilde{p} = \left(\ln (x_1/x_0)\right)^{-1}$. 
If the distribution is very stretched reasonable approximations are  
\be{0}
s &\approx& 2\tilde p
\\
p &\approx& - \tilde p\ln \tilde p
\ee
For the VRH calculation
\be{0}
x_{\omega} =
x_1\left(\frac{x_0}{x_1}\right)^{1\big/{\varrho_{\tbox{E}}\omega}} 
\approx
\frac{\langle\langle x\rangle\rangle_a}{\tilde p}
\exp\left(-\frac{1}{\tilde p \varrho_{\tbox{E}}\omega}\right)
\ee
For a rectangular $\tilde{F}(\omega)$ the VRH optimization 
is trivial and gives $\omega \approx x_{\omega_c}$, leading to
\be{0}
g_{\tbox{SLRT}} \ \ \approx \ \ 
\frac{1}{\tilde p} \ 
\exp\left[-\frac{1}{\tilde p \varrho_{\tbox{E}}\omega_c } \right]  
\ee
For an exponential $\tilde{F}(\omega)$ the VRH optimization gives
\be{0}
g_{\tbox{SLRT}} \ \ \approx \ \ 
\frac{1}{\tilde p} \exp \left[
-2\left({\frac{1}{\tilde p \varrho_{\tbox{E}}\omega_c}}\right)^{1/2}\right]
\ee
which is the same as in the traditional VRH optimization.

%%%%%%%%%%%%%%%%%%%%%%%%%%%%%%%%%%%%%%%
\section{The log-normal ensemble}
\label{app:log-normal}

The cumulative distribution function that corresponds to Eq.~(\ref{e17}) is
\be{0}
\mbox{Prob}(X < x) = \frac{1}{2}+\frac{1}{2}
\mathrm{erf}\left[\frac{\ln(x)-\mu}{\sigma\sqrt{2}}\right]
\ee
The algebraic, geometric and harmonic averages are
\be{0}
\langle\langle x\rangle\rangle_a &=& \eexp{\mu + \sigma^2/2} \\
\langle\langle x\rangle\rangle_g &=& \eexp{\mu} \\
\langle\langle x\rangle\rangle_h &=& \eexp{\mu - \sigma^2/2}
\ee
The sparsity parameters are  
\be{0}
s &=& q^2
\\
p &=& \frac{1}{2} \mbox{erfc}\left(\frac{\sigma}{2\sqrt{2}} \right) 
\\
q &=& \eexp{- \sigma^2/2} 
\ee
The VRH estimate is
\be{0}
x_\omega &=& \exp\left[\mu + \sigma\sqrt{2}
\ \mbox{erfinv}\left(1 - \frac{2}{\varrho_{\tbox{E}}\omega}\right)\right] \\
&\approx &
\langle \langle x \rangle\rangle_g 
\exp\!\left[
\sqrt{\ln \left(\frac{1}{q}\right)^2
\left(\ln\frac{\varrho_{\tbox{E}}^2\omega^2}{2\pi} -
\ln\ln\frac{\varrho_{\tbox{E}}^2\omega^2}{2\pi} \right)} 
\right] \nonumber
\ee
For a rectangular $\tilde{F}(\omega)$ the VRH optimization 
is trivial and gives $\omega \approx x_{\omega_c}$, leading to
\be{0}
g_{\tbox{SLRT}} \ \ \approx \ \ 
q \ \exp\left[2 \sqrt{-\ln q \ln (\varrho_{\tbox{E}}\omega_c)} \right] 
\ee
For an exponential $\tilde{F}(\omega)$ the VRH optimization gives 
\be{0}
g_{\tbox{SLRT}}  &\approx& 
q \ 
\exp\left[\sqrt{-\ln q \ln \frac{-\varrho_{\tbox{E}}^2 \omega_c^2 \ln q}{\pi}} 
- \sqrt{\frac{\sqrt{-4\pi\ln q}}{\varrho_{\tbox{E}}\omega_c}}\right]  
\nonumber \\
&\approx& q \ \exp\left[ \sqrt{-2\ln q \ln (\varrho_{\tbox{E}}\omega_c)}
\right] 
\ee
Due to the minority dominance the functional form 
is more robust compared with the log-box case.

%%%%%%%%%%%%%%%%%%%%%%%%%%%%%%%%%%%%%%%%%%%%%%%%%%%%%%%%%%%%%%%%%%%%%%%%%%%%%%%%%%%%%%%%%%%
%%%%%%%%%%%%%%%%%%%%%%%%%%%%%%%%%%%%%%%%%%%%%%%%%%%%%%%%%%%%%%%%%%%%%%%%%%%%%%%%%%%%%%%%%%%
%%%%%%%%%%%%%%%%%%%%%%%%%%%%%%%%%%%%%%%%%%%%%%%%%%%%%%%%%%%%%%%%%%%%%%%%%%%%%%%%%%%%%%%%%%%
%%%%%%%%%%%%%%%%%%%%%%%%%%%%%%%%%%%%%%%%%%%%%%%%%%%%%%%%%%%%%%%%%%%%%%%%%%%%%%%%%%%%%%%%%%%

%%%%%%%%%%%%%%%%%%%%%%%%%%%%%%%%%%%%%%%%%%%%%%%%%%%%%%%%%%%%%%%%
%%%%%%%%%%%%%%%%%%%%%%%%%%%%%%%%%%%%%%%%%%%%%%%%%%%%%%%%%%%%%%%%
%%%%%%%%%%%%%%%%%%%%%%%%%%%%%%%%%%%%%%%%%%%%%%%%%%%%%%%%%%%%%%%%
%%%%%%%%%%%%%%%%%%%%%%%%%%%%%%%%%%%%%%%%%%%%%%%%%%%%%%%%%%%%%%%%

%FIGURES

\ \\ \ \\ 

%%%%%%%%%%%%%%%%%%%%%%%%%%%%%%

\clearpage

%%%%%%%%%%%%%%%%%%%%%%%%%%%%%%

%%%%%%%%%%%%%%%%%%%%%%%%%%%%%%

%%%%%%%%%%%%%%%%%%%%%%%%%%%%%%

%%%%%%%%%%%%%%%%%%%%%%%%%%%%%%

%%%%%%%%%%%%%%%%%%%%%%%%%%%%%%

%%%%%%%%%%%%%%%%%%%%%%%%%%%%%%

%%%%%%%%%%%%%%%%%%%%%%%%%%%%%%

\clearpage

%%%%%%%%%%%%%%%%%%%%%%%%%%%%%%%%%%%%%%%%%%%%%%%%%%%%%%%%%%%%%%%%
%%%%%%%%%%%%%%%%%%%%%%%%%%%%%%%%%%%%%%%%%%%%%%%%%%%%%%%%%%%%%%%%
\end{document}